\documentclass[12pt,preprint,a4paper]{aastex} %MNRAS

\usepackage{amsmath}                % American Mathematical Society package
\usepackage{amsfonts}               % American Mathematical Society fonts
\usepackage{amssymb}                % American Mathematical Society symbol
\usepackage{epsfig}                 % EPS figures
\usepackage{url}                       % url links
\usepackage{times}

\newcommand{\s}{{~\rm s}}
\newcommand{\km}{{~\rm km}}

\newcommand{\erg}{{~\rm erg}}
\newcommand{\yr}{{~\rm yr}}

\newcommand{\AU}{{~\rm AU}}

\begin{document}

\title{A PRE-EXPLOSION OPTICAL TRANSIENT EVENT FROM A WHITE DWARF MERGER WITH A GIANT SUPERNOVA PROGENITOR}

\author{Efrat Sabach\altaffilmark{1} and Noam Soker\altaffilmark{1}}

\altaffiltext{1}{Department of Physics, Technion -- Israel Institute of Technology, Haifa 32000,
Israel; efrats@physics.technion.ac.il; soker@physics.technion.ac.il}

\begin{abstract}
We examine rare evolutionary routes of binary systems where the initially more massive primary star of $M_{1,0}\simeq 5.5-8.5M_\odot$,
forms a white dwarf (WD), while the secondary star of $4 M_\odot\la M_{2,0} < M_{1,0}$, accretes mass from the evolved primary
and later terminates as a core collapse supernova (CCSN).
In such a {\it WD-NS reverse evolution} a neutron star (NS) 
or a potential NS-progenitor massive core is formed after the WD.
These SN explosions are likely to be preceded by strong interaction of the WD with the giant secondary's core,
leading to an Intermediate-Luminosity Optical Transient (ILOT; Red Transient; Red Nova) event, weeks to years before the explosion.
The common envelope phase of the WD and the giant ends with a merger that forms an ILOT,
or an envelope ejection that leads, after a CCSN of the giant's core, to a NS-NS or WD-NS surviving binary.
The WD could suffer a thermonuclear explosion, that might be observed as a Type Ia SN.
Most of these CCSN and thermonuclear explosions will be peculiar.
We calculate the stellar evolution of representative cases using Modules
for Experiments in Stellar Astrophysics (MESA).
The occurrence rate of these systems is $\sim 3-5$ percent of that of CCSNe.
\end{abstract}


% ====================
\section{INTRODUCTION}
\label{sec:introduction}
% ====================

With better sky coverage more and more rare transient events with peak luminosity between those of novae and supernovae (SNe) are detected
(e.g., \citealt{Mould1990, Rau2007, Ofek2008, Berger2009, Botticella2009, KulkarniKasliwal2009, Prieto2009, Smith2009, Mason2010, Pastorello2010, Berger2011, Kasliwal2011, Tylendaetal2013}).
We refer to them as ILOTs, for Intermediate-Luminosity Optical Transients, although  Red Novae and Red Transients are also in use.
These rare eruptions can typically last weeks to several years.
The pre-outburst objects of some of the ILOTs, e.g., NGC~300 OT2008-1 (NGC~300OT; \citealt{Bond2009}), are asymptotic giant branch (AGB) or extreme-AGB stars.
There are single star models (e.g., \citealt{Kochanek2011}) and binary stellar models
\citep{Kashietal2010, SokerKashi2013} for ILOT events harboring AGB stars.
We consider the binary model ILOTs, and in the present paper study a specific binary evolutionary channel.

In the binary paradigm ILOTs can be powered from merger of a main sequence (MS) star (or slightly evolved off the MS) with another MS star,
as in V838~Mon \citep{SokerTylenda2003} and V1309~Sco \citep{Tylendaetal2011}, and from a MS star accreting mass from an evolved star \citep{KashiSoker2010b, Kashietal2013}.
\cite{Ivanovaetal2013a} suggested that after merger in the formation of a common envelope (CE) phase, ILOTs can be controlled by recombination and powered by the energy released from the recombination of the ejected gas. For the ILOTs considered here, that result from merger of a white dwarf (WD) and a core, \textit{WD-core merger}, the recombination energy is negligible.

ILOT can be also powered at the termination of the CE phase by the secondary merging with the core.
\cite{Tylendaetal2013} proposed that the ILOT (red transient) OGLE-2002-BLG-360 was powered by the collision of a secondary with the core of an evolved star.
In some cases an ILOT event can precede a core collapse SN (CCSN) event, such as in SN 2010mc \citep{Ofeketal2013}.
SN~2010mc is a Type IIn CCSN where the ejecta is thought to interact with close circumstellar matter (CSM).
\cite{Soker2013b} speculated that the pre-explosion outburst (PEO) was energized by mass accretion onto an O main-sequence stellar secondary.
More generally, some ultraluminous CCSNe experience an extreme mass loss episode $\sim 1-100 \yr$ before explosion (e.g., \citealt{Ofeketal2007}, \citealt{Smithetal2007}, \citealt{Chomiuk2011} and a review by \citealt{Gal-Yam2012}, and references therein; Also, see discussion in \citealt{Chevalier2012}).
\cite{Chevalier2012} and \cite{Soker2013a} suggested that this coincidence can be caused by a secondary that spirals-in inside the envelope and collides with the core.
In the scenario of \cite{Chevalier2012} the secondary is a neutron star (NS) or a black hole (BH) that can accrete mass from the
primary and launch jets. \cite{Soker2013a} was considering a MS secondary.

In the present paper we consider rare cases, termed {\it WD-NS reverse evolution}, where the compact companion to the progenitor of a CCSN is a white dwarf (WD), as opposed to the classical approach of binary systems where the NS is formed first, termed {\it NS-WD}.
 A mass transfer that leads the secondary (initially less massive) star to explode as a CCSN and form a WD-NS system with a NS younger than the WD, was mentioned before
(e.g.,  \citealt{TutukovYungelSon1993, PortegiesZwartVerbunt1996, vanKerkwijkKulkarni1999, PortegiesZwartYungelson1999, TaurisSennels2000, Brownetal2001, Nelemansetal2001, Daviesetal2002, Kimetal2003, Kalogeraetal2005, Churchetal2006, vanHaaftenetal2013}),
e.g. to explain the presence of a massive WD in the binary radio pulsars PSR B2303+46 \citep{ PortegiesZwartYungelson1999, vanKerkwijkKulkarni1999,  TaurisSennels2000,  Brownetal2001, Daviesetal2002, Kalogeraetal2005, Churchetal2006} and PSR J1141-6545 \citep{TaurisSennels2000, Brownetal2001, Daviesetal2002, Kalogeraetal2005, Churchetal2006}. As both systems are in an eccentric orbit the NS-WD evolution is ruled out since it leads to a circularized orbit.
While these previous works focus mainly on an outcome of a binary containing a WD and a NS, our present work emphasizes other outcomes of the reverse evolution, in particular WD-core merger and the possibility of an ILOT event.
We also note that \cite{Sipioretal2004} considered a mass transfer process in more massive binary systems that leads to a reversal of the end states, resulting in a NS that forms before a black hole.

In section \ref{sec:evolution} we list the evolutionary routes of the WD-NS reverse evolution considered in the present study and also estimate the Galactic birthrate of such systems (section \ref{sec:BR}). In section \ref{sec:CE} we discuss the possible outcomes of the CE evolution. In section \ref{sec:postCE} we discuss the observational consequences of some of the outcomes. Our summary is in section \ref{sec:Summary}.

% ====================
\section{PRE-COMMON ENVELOPE EVOLUTION}
\label{sec:evolution}
% ====================

\begin{figure}[ht]
\centering
\includegraphics[width=140mm]{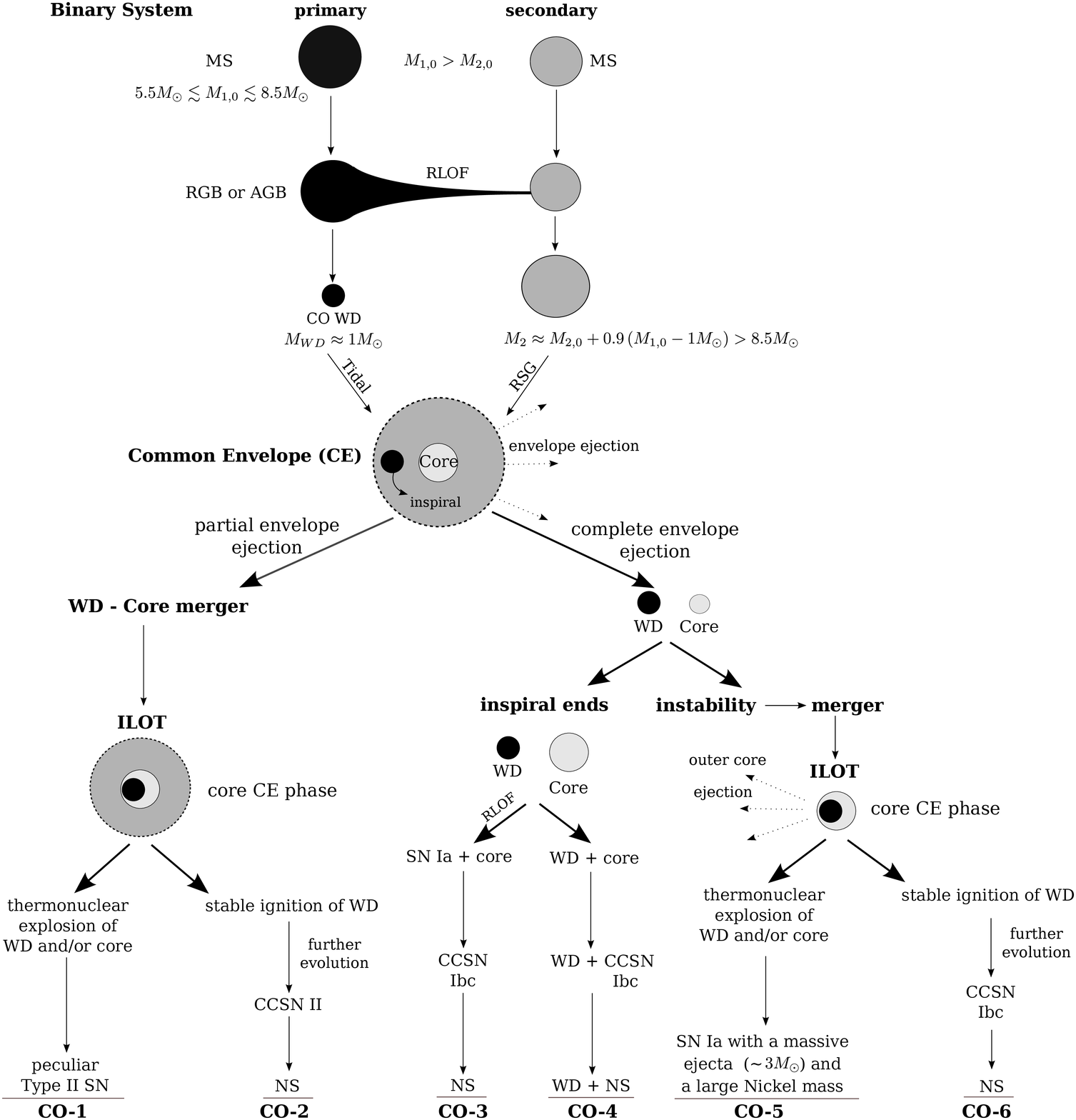}
\caption{\footnotesize Schematic evolutionary paths of the binary systems studied here, where the remnant of the primary star (black), of mass $M_{1,0}\simeq 5.5-8.5M_\odot$, is a CO white dwarf (WD). The different evolutionary routes for a CO WD are numbered in the bottom line.
We study systems where the primary experiences a Roche lobe overflow (RLOF) when it reaches either the RGB or AGB phase (2$^{\rm nd}$ stage in the figure). 
If the primary experiences RLOF before it develops a CO core as a giant, an intermediate stage will take place (before the 3$^{\rm rd}$ stage): the giant primary loses its H-rich envelope, shrinks, and the He core continues to evolve and form a CO core after the RLOF ceased.
We consider only systems where even if mass transfer starts when the primary is on the RGB sufficient envelope is left for it to form a CO WD.
The mass transfer brings the secondary mass to $M_2 > 8.5 M_\odot$, and leaves a WD remnant from the primary star.
As the secondary star (gray) evolves to a red supergiant (RSG), tidal forces cause the WD to lose angular momentum and spiral-in towards the core of the secondary RSG (4$^{\rm th}$ stage). The CE can terminate in one of three routes.
(a) The WD merges with the core before the entire envelope is ejected. Any explosion that occurs in one of the cases indicated will lead to a type II SN. In some of the cases the explosion will lead to a usual CCSN process, and in others a thermonuclear explosion will lead to a peculiar type II SN.
(b) The entire envelope is lost and the WD ends outside the core of the RSG. This route will lead to either a type Ia supernova (SN Ia) + core collapse SN (CCSN), or a WD + CCSN. Both CCSN will be either type Ib or Ic (Type Ibc in short).
(c) The entire envelope is ejected and the system suffers a dynamical instability, e.g., Darwin instability, such that a WD-core merger occurs.
A second CE phase commences. This liberates huge amounts of gravitational energy that can manifest as an intermediate luminosity optical transient (ILOT).
This route with a CO WD will lead to either a thermonuclear explosion leaving nothing behind, or a CCSN resulting in a neutron star.}
\label{fig:mech1}
\end{figure}

\begin{figure}[ht]
\centering
\includegraphics[width=165mm]{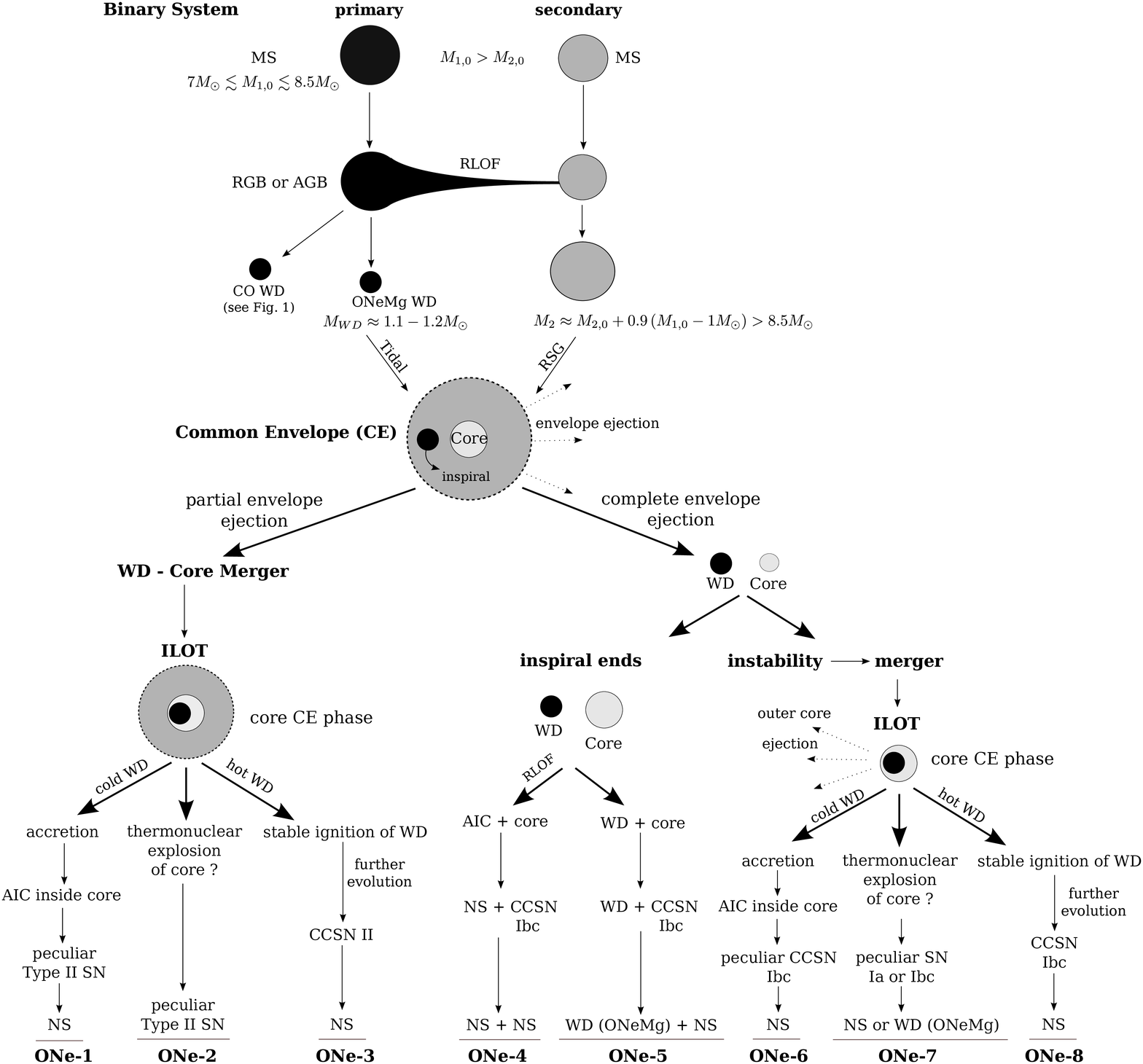}
\caption{\footnotesize Like Fig. \ref{fig:mech1} but for the case where the primary star, of mass $M_0=7-8.5M_\odot$, leaves an ONeMg WD remnant (\citealt{Kroupa1993}).
With an ONeMg WD in some cases the outcome might be accretion induced collapse (AIC) or a CCSN instead of a thermonuclear explosion that occurs for a CO WD. The different evolutionary routes for a ONeMg WD are numbered in the bottom line. }
\label{fig:mech2}
\end{figure}

We examine the evolution of two massive stars in a binary system with a total mass of $\sim 10-15M_\odot$
as schematically presented in Figs. \ref{fig:mech1} and \ref{fig:mech2}.
Each one of the original stars by itself will end up in a white dwarf (WD), but due to mass transfer the secondary might become
massive enough to be considered a progenitor of a CCSN.
The binary evolution might be accompanied by an ILOT event.
As seen in these figures we consider many evolutionary routes where a WD is formed before either a NS, which we term WD-NS reverse evolution, or a massive core that is a potential progenitor of a NS; we will refer in short to the second case as WD-NS reverse evolution as well. 
Several papers other than those listed in section \ref{sec:introduction} include charts of CE evolution and NS formation that show the rich variety of evolutionary routes involving a CE phase, e.g., \cite{IbenTutukov1984},
\cite{Hanetal1995}, \cite{Chevalier2012}, \cite{Toonenetal2012} and \cite{Dall'OssoPiran2013};
another chart with more examples and a thorough review of the CE evolution can be found in \cite{{Ivanovaetal2013b}}.
However, these papers did not emphasize the WD-NS reverse evolution accompanied by a possible merger event, and did not consider the possibility of an ILOT event in a short or extended time before the SN explosion.

% ============
\subsection{Binary evolution}
\label{sec:Evolutionary_model}
% ============

\begin{figure}[ht]
\centering
\includegraphics[width=81.5mm]{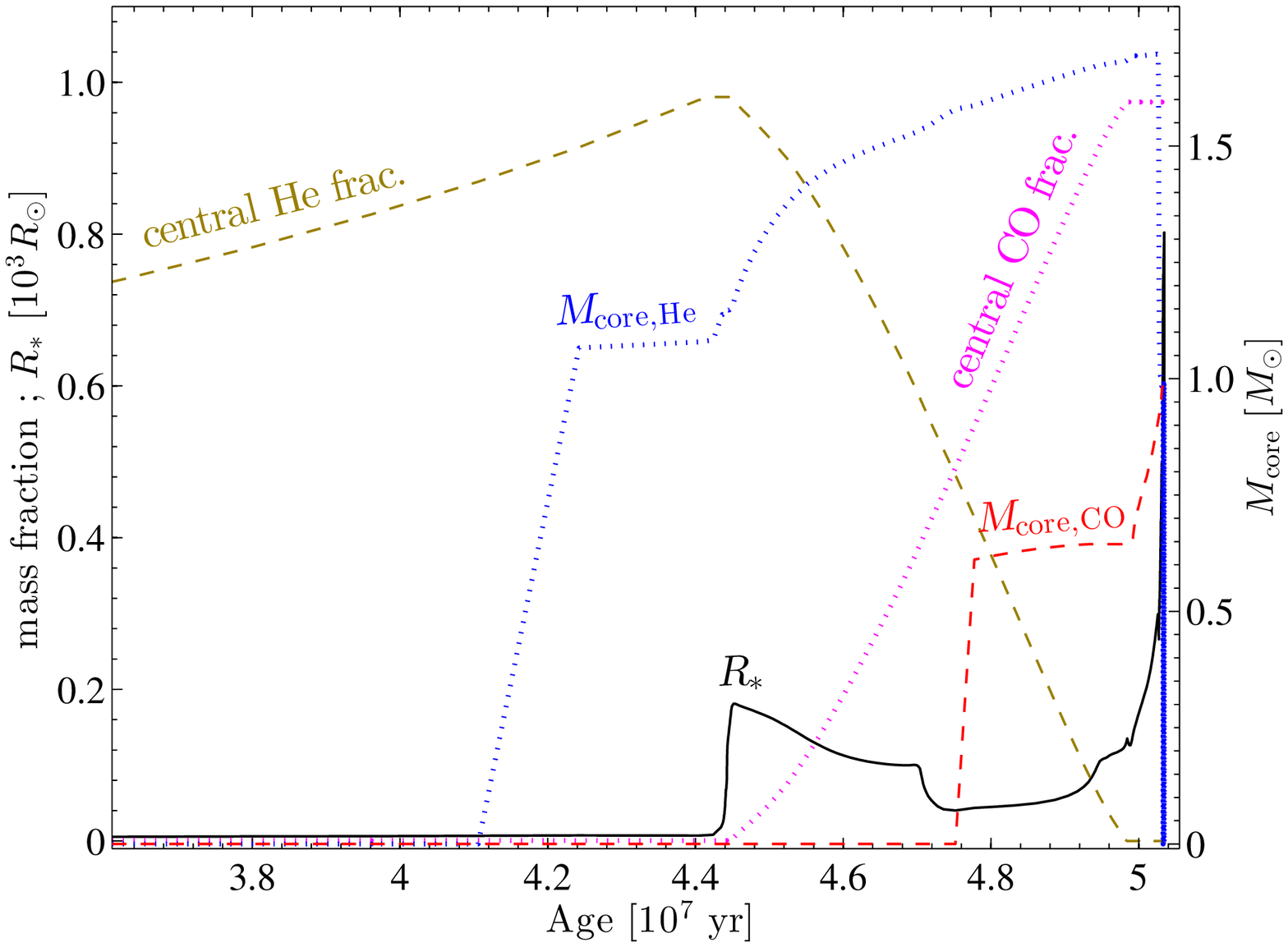}
\hspace*{0.4cm}
\includegraphics[width=75mm]{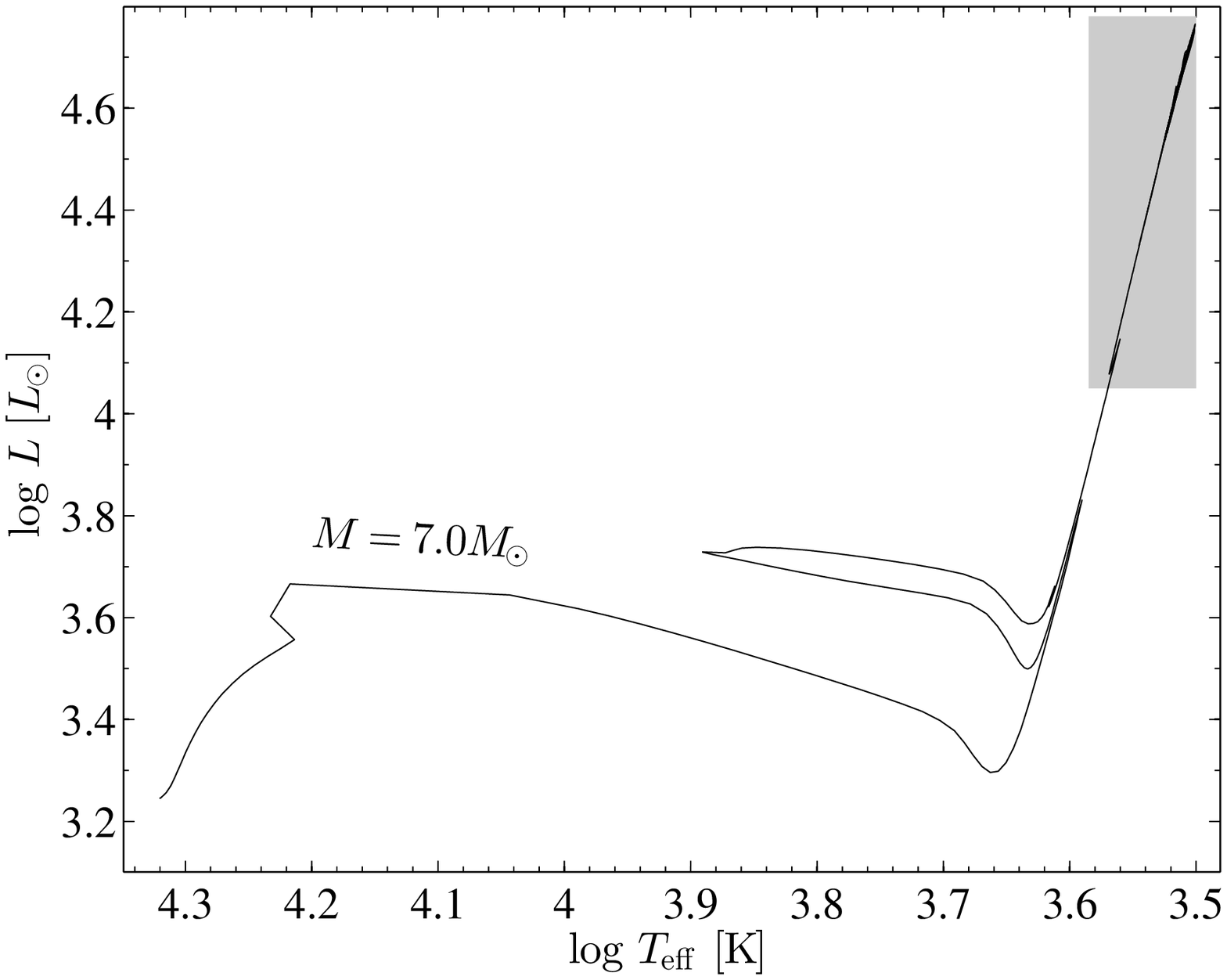}
\caption{Evolution of a $M_{1,0}=7 M_\odot$ primary star calculated with MESA (\citealt{Paxton2011}), until the second expansion phase.
Left panel: The radius, mass fractions of He and C+O in the core (left axis) and mass of the He core and C+O core (right axis) are plotted
as function of time. The core mass fraction of He and CO is defined close to the center of the star.
The He (CO) core mass is defined as the mass at the point where the He (CO) fraction is $0.5$.
Taking a mass fraction of 0.9 changes the core mass by a negligible amount.
There are two expansion phases where RLOF could potentially occur. Right panel: The evolution on the HR diagram. Gray rectangle marks the phase of thermal pulsations. }
\label{fig:7MS}
\end{figure}

We calculate stellar evolution for non-rotating stars with solar metallicity (Z=0.02) from the ZAMS, with initial mass of $\simeq 4-8.5M_\odot$ for the original stars and $ M_{2}\simeq 8.5-14M_\odot$ for the post-mass transfer secondary star. In all calculations we use the Modules for Experiments in Stellar Astrophysics (MESA), version 4798 (\citealt{Paxton2011}).
For a primary of mass $M_{1,0}\simeq 5.5-8.5M_\odot$, where $M_{1,0}$ is the initial mass of the primary star, two rapid rises in the radius are expected, one when the core is mainly made out of helium and the second when the star has a CO core.
These two expansion phases are presented for a $7 M_\odot$ stellar model in Fig. \ref{fig:7MS}.
Despite the limited temporal resolution of the evolution presented on the HR diagram, it is very similar to that presented by \cite{Ekstrometal2012} for their $7 M_\odot$ star.
We study binary systems where the orbital separation is such that a Roche Lobe overflow (RLOF) occurs either when the primary star has a He core, or during its second expansion phase when it has a massive CO core. In the earlier case the primary transfers its H-rich envelope, shrinks, mass transfer ceases, and the He core continues to evolve to form a CO WD. Here we do not treat the mass transfer phase, but only assume an ad-hoc mass transfer rate allowing the accreting star to dynamically re-adjust, hence preventing a common envelope (CE)
during the mass transfer phase \citep{Churchetal2006}.
The RLOF commences before a significant amount of mass is removed from the primary by its wind.
At most about several $0.1 M_\odot$ of mass is carried by the primary wind and possible jets from the accreting secondary star.
Moreover, we also consider only massive secondary stars, $M_{2,0} \ga 4M_\odot$, where  $M_{2,0}$ is the initial mass of the secondary star, such that the secondary (a) can accrete at a high rate, several $M_\odot$ within $\sim 10^3-10^4 \yr$, from the mass-losing primary star, and (b) brings the primary envelope to synchronization and a CE is avoided at this stage.
 This mass limit of $M_{2,0} \ga 4 M_\odot$ is very similar to that used by \cite{TaurisSennels2000}.
After mass transfer ends the secondary has a mass of $M_2 > 8.5 M_\odot$, where  $M_2$ is the mass of the post-accretion secondary star, so that it can be considered as a CCSN progenitor (\citealt{Langer2012}).

We make the same assumptions as in the population synthesis of \cite{IlkovSoker2013}. In particular  
that (a) $\eta=0.9$ of the mass lost by the primary is accreted by the secondary, and that
(b) the secondary is massive enough, $q=M_{2,0}/M_{1,0}\gtrsim 0.45$, to bring the primary envelope to synchronization and prevent a CE phase during the first mass transfer episode (from the primary to the secondary). This is similar to the treatment of \cite{TaurisSennels2000}.
We note the following regarding these assumptions: 
(1) An accretion fraction of $\eta=0.9$ can be achieved for example if the mass is lost mainly by jets blown by the accreting companion, then a mass loss fraction of $1-\eta=0.1$ is typical. 
 Another way to get $\eta \simeq 0.9$ is that the mass transfer is conservative in the RLOF until the primary loses its H-rich envelope and shrinks to form a He star, and then loses the rest of the envelope in a wind \citep{TaurisSennels2000}.
(2) Such jets and radiation from the accretion disk remove most of the accreted energy, such that the secondary can accrete lots of mass within a relatively short time without expanding much, e.g., as the secondary in the Great Eruption of $\eta$ Carinae \citep{KashiSoker2010a}.
(3) After the mass of the secondary becomes larger than the primary mass, mass transfer increases the orbital separation, helping in avoiding the CE phase.   
These assumptions are in dispute, but bring the number of Type Ia progenitors to a much better agreement  with observations \citep{IlkovSoker2013}.
In the calculations of stellar evolution we use conservative mass transfer ($\eta=1$) in the discussions, just for convenience.

% FFFFFFFFFFFFFFFFFFFFFFFFFFFFFFFFFFFFFFFFFFFFFFFFFF
% FFFFFFFFFFFFFFFFFFFFFFFFFFFFFFFFFFFFFFFFFFFFFFFFFF
% FFFFFFFFFFFFFFFFFFFFFFFFFFFFFFFFFFFFFFFFFFFFFFFFFF
% FFFFFFFFFFFFFFFFFFFFFFFFFFFFFFFFFFFFFFFFFFFFFFFFFF
% FFFFFFFFFFFFFFFFFFFFFFFFFFFFFFFFFFFFFFFFFFFFFFFFFF
% FFFFFFFFFFFFFFFFFFFFFFFFFFFFFFFFFFFFFFFFFFFFFFFFFF

We assume that most of the envelope is transferred to the secondary star, and the rest is blown away by winds.
The leftover from the primary  is a WD of $M_{\rm WD} \simeq 0.8-1.2 M_\odot$. The WD can be either a CO WD or an ONeMg WD. The later case might
occur when the primary initial mass is $7M_\odot\la M_{1,0} \la 8.5 M_\odot$.
The mass of the secondary is now $M_2 \simeq M_{2,0}+0.9 \left( M_{1,0}- 1 M_\odot\right)>8.5M_\odot$ or somewhat lower, where we took an accretion fraction of $\sim 0.9$. This also sets the lower boundary on the initial mass of the primary to be $\simeq 5.5M_\odot$. We included in our calculations the Reimers wind scheme on the RGB and the Blocker wind scheme on the AGB; no post AGB wind was included.
Once the post accretion secondary evolves and expands as a giant, a CE stage begins because the WD cannot bring the envelope to synchronization and tidal
forces cause the WD to spiral-in into the secondary inflated envelope.

\begin{figure}[ht]
\centering
\includegraphics[width=112mm]{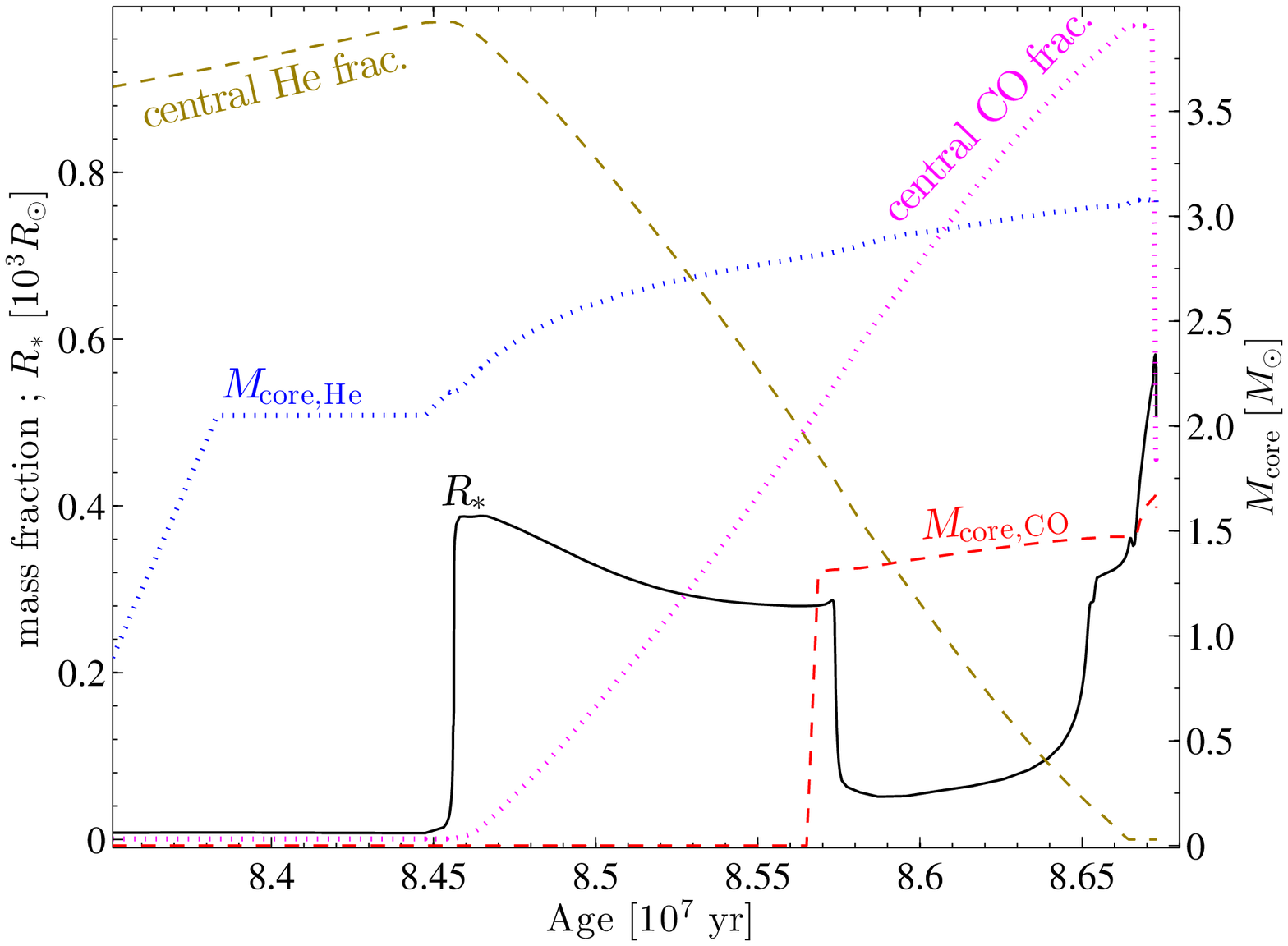}
\caption{Post accretion evolution of a $10M_\odot$ secondary star. The CE phase will occur either in the first jump in radius, when the giant develops a He core, or at the second rise, when the giant has a CO core.  }
\label{fig:ev_10}
\end{figure}

\begin{figure}[ht]
\centering
\includegraphics[width=112mm]{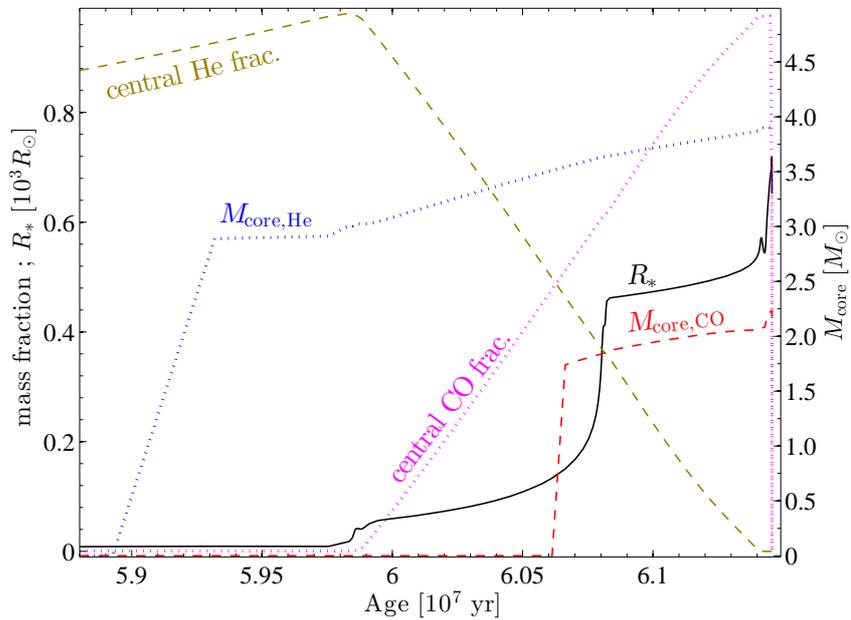}
\caption{Like Fig. \ref{fig:ev_10}, but for the post accretion of the $12M_\odot$ secondary star. The CE phase will occur either in the first significant jump in radius, when the giant develops a HeCO core, or at the second rise, when the giant has a CO core.  }
\label{fig:ev_12}
\end{figure}

\begin{figure}[ht]
\centering
\includegraphics[width=79mm]{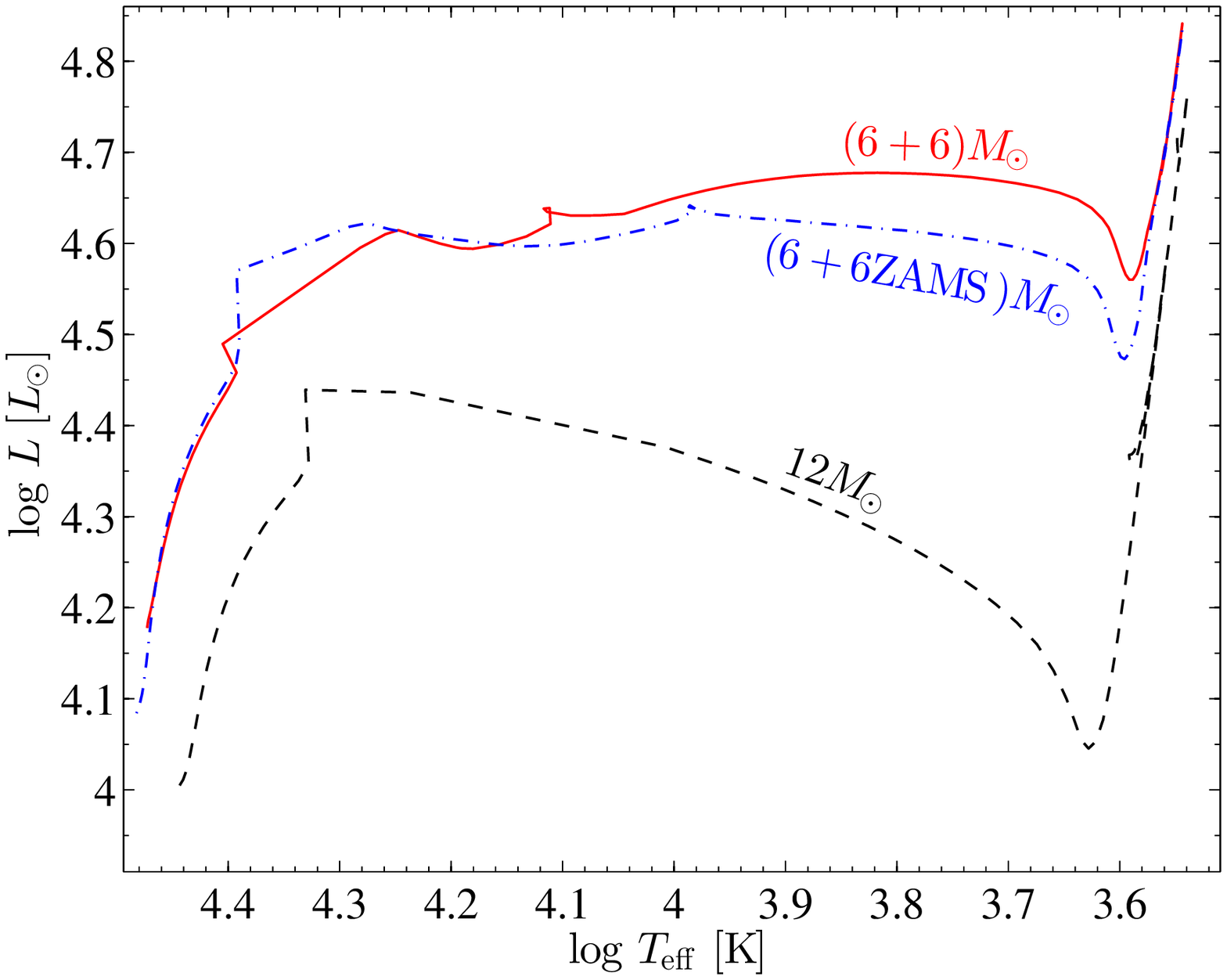}
\hspace*{0.4cm}
\includegraphics[width=79mm]{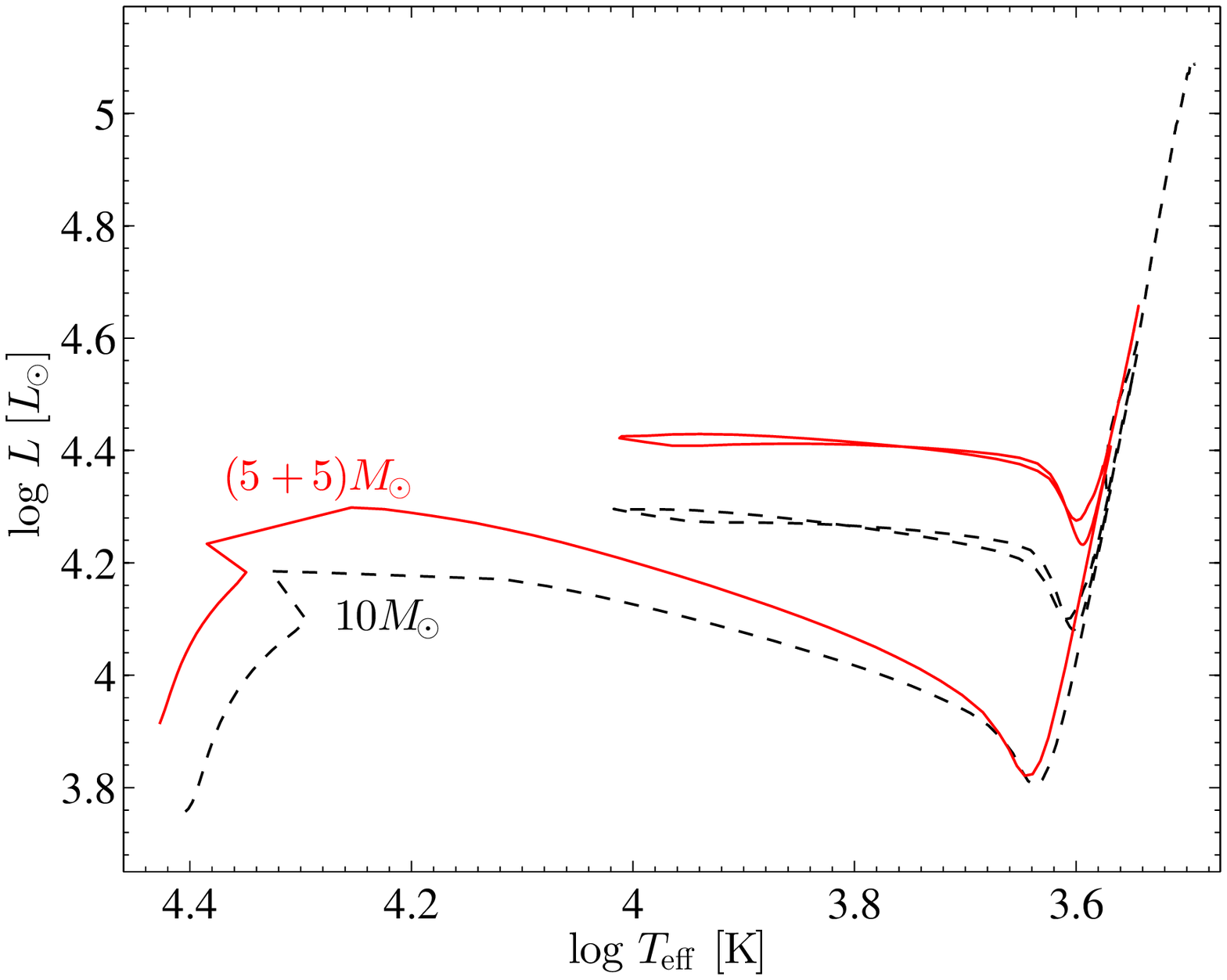}
\caption{Evolution of the two post-accretion models on the HR diagram.
Left panel: Red (upper) solid line depicts the evolution of the post-accretion $12 M_\odot$ star that we use in the present study.
For comparison the evolution of a ZAMS star of $12 M_\odot$ and solar composition is shown by the dashed-black (lower) line. The differences in the evolution result from the
He-enrichment of the post-accretion model. To evaluate the role of He-enrichment in the accreted envelope, the dashed-dotted blue line shows the evolution
of a post-accretion $12 M_\odot$ model where only the accreted gas is He-enriched.
From the comparison of the three models it is evident that the He-enrichment of the accreted gas plays a
larger role than the He-enriched core in determining the post-accretion evolution of the secondary star.
One caveat here is the that the MESA mixing prescription of the accreted gas into the star might be too slow (see text).
Right panel: The same as in the left panel but for a $10 M_\odot$ post-accretion model and a $10 M_\odot$ star from ZAMS.}
\label{fig:HR_comp}
\end{figure}

We use MESA to follow the evolution of the secondary star from the beginning of the MS, through accretion, relaxation to the MS after accretion,
and then evolution to the formation of a CO core and even beyond, as illustrated in Figures \ref{fig:ev_10} and \ref{fig:ev_12}, for a $10M_\odot$ secondary and for a $12M_\odot$ secondary, respectively. The relaxation time from the end of accretion to the MS is about equal to the thermal time
of the post-accretion star on the MS.
The accretion onto the secondary star starts when the primary star expands to a giant, either as a RGB or an AGB star. The transferred mass is enriched with helium due to dredge up.
In addition, by that time the secondary core is helium enriched.
After accretion the secondary is He-enriched (both the core and the envelope) and therefore the post-accretion evolution is not
identical to that of a star that starts the MS with the same mass and with solar composition.
This is demonstrated in Fig. \ref{fig:HR_comp}, where we compare the evolution of our post-accretion secondary to the evolution of a ZAMS star with the same mass.
To separate the influence of the He-enriched secondary core and the He-enriched accreted mass, for the case of a $12M_\odot$ star we run a third model with accretion of a
He-enriched gas onto a ZAMS secondary star. From the comparison of the three models it is evident that the He-enrichment of the accreted gas plays a
larger role than the He-enriched core in determining the post-accretion evolution of the secondary star.
Since the accreted matter is He-inriched compared to the secondary pre-accretion envelope, the invert composition gradient will lead to thermohaline mixing on a time-scale of the order of the thermal time-scale of the star \citep{Churchetal2006}.
MESA includes thermohaline mixing in a diffusion approximation that is much slower than the thermal readjustment process. We don't expect the inefficient mixing in MESA to influence our results and conclusions as we are more sensitive to the core evolution than the envelope evolution.

The post-accretion stellar models with $ 8.5M_\odot < M_2 \la 14 M_\odot$ will be used next to study the possible outcomes of their
interaction with the WD descendant of the primary star.
The post-accretion secondary star expands to a radius comparable to the maximum radius attained by the original primary star, but now the orbital angular momentum is too low to bring the secondary envelope to synchronization and a CE is inevitable.
As seen in Figures \ref{fig:ev_10} and \ref{fig:ev_12}, the post-accretion secondary has two possible expansions,
the first when it obtained a He core (or HeCO core for the $12M_\odot$ star),
and the other when there is an inner CO core surrounded by a He shell (which we refer to as a CO core).
 Most likely the WD enters the envelope during the first expansion phase, when the core is He for the $10 M_\odot$ star,
or HeCO for the $12 M_\odot$ star.
The reason is that the evolution time during the first expansion is much longer than during the second expansion phase, such that tidal interaction
has time to cause the WD to spiral-in from a distance of $a_{\rm max} \sim 5 R_g$, where $R_g$ is the maximum radius during the first expansion phase.
These are rare types of systems where a WD orbits a star that potentially can form a NS via a CCSN. Namely, the WD is formed {\it before} the NS, as we term this WD-NS reverse evolution.
% FFFFFFFFFFFFFFFFFFFFFFFFFFFFFFFFFFFFFFFFFFFFFFFFFF
% FFFFFFFFFFFFFFFFFFFFFFFFFFFFFFFFFFFFFFFFFFFFFFFFFF
% FFFFFFFFFFFFFFFFFFFFFFFFFFFFFFFFFFFFFFFFFFFFFFFFFF
% FFFFFFFFFFFFFFFFFFFFFFFFFFFFFFFFFFFFFFFFFFFFFFFFFF
% FFFFFFFFFFFFFFFFFFFFFFFFFFFFFFFFFFFFFFFFFFFFFFFFFF
% FFFFFFFFFFFFFFFFFFFFFFFFFFFFFFFFFFFFFFFFFFFFFFFFFF

% ====================
\subsection{Birthrate estimation}
\label{sec:BR}
% ====================

To estimate the number of the studied systems relative to all potential progenitor of CCSNe we proceed as follows.
For the initial mass function of the relevant primary we take \citep{Kroupa1993}
\begin{equation}
\frac {d N}{d M} = A M^{-2.7}, \quad {\rm for} \quad 1.0M_\odot < M ,
\label{eq:IMF1}
\end{equation}
where A is a constant.
For the progenitor of CCSNe we take all stars of $M>8.5M_\odot$, for which integration gives $N_{\rm CCSN} = 0.015 A$.
For the systems studied here we demand that the initial secondary stellar mass must be $M_{1,0} > M_{2,0} \gtrsim 4 M_\odot$, and that the primary initial mass must be
$M_{1,0} \gtrsim 5.5 M_\odot$ in order to allow $M_2>8.5M_\odot$.
We also assume a flat mass ratio distribution. We take $f_{b} \simeq 60 \%$ of primary B-type stars to be in binary systems \citep{Raghavanetal2010}.

The number of relevant binary systems by mass is given by
\begin{equation}
N_{b} \simeq \int   \frac {d N_1}{d M_{1,0}} \left( \frac{M_{1,0}-M_{2,0,{\rm{min}}}}{M_{1,0}} \right)dM_{1,0};
\label{eq:Pr1}
\end{equation}
Where $M_{2,0,{\rm{min}}}$ is the minimum allowed secondary mass, and it depends on the Primary mass (see below).
Taking into account the mass accretion fraction, $\eta=0.9$, for any initial mass of the primary the post accretion secondary must satisfy
$M_2 \simeq M_{2,0}+0.9 \left( M_{1,0}- 1 M_\odot\right)>8.5M_\odot$  in order to be considered a CCSN progenitor.
Since we take the lower boundary on the initial mass of the companion to be $4M_\odot$ any primary star with a mass within the range $6M_\odot\lesssim M_{1,0}<8.5M_\odot$ would allow the mass of the post accretion companion to be $M_2>8.5M_\odot$.
Namely, the minimum allowed mass for the secondary is $M_{\rm 2,0,min} = 4M_\odot$ for $M_{1,0}\ga 6.5 M_\odot$.
For the lower mass range of primary stars $5.5M_\odot\lesssim M_{1,0}<6M_\odot$ the minimum allowed initial secondary mass has to satisfy $M_{2,0,{\rm{min}}}< 9.4M_\odot-0.9M_{1,0}$ so that $M_2>8.5M_\odot$ holds.
Consequently we divide the integration in equation (\ref{eq:Pr1}) to two parts
\begin{equation}
N_{b} \simeq \int^{6M_\odot}_{5.5M_\odot}   \frac {d N_1}{d M_{1,0}} \left( \frac{1.9M_{1,0}-9.4 M_\odot}{M_{1,0}} \right)dM_{1,0}
+ \int^{8.5M_\odot}_{6M_\odot}   \frac {d N_1}{d M_{1,0}} \left( \frac{M_{1,0}-4 M_\odot}{M_{1,0}} \right)dM_{1,0}
\label{eq:Pr2}
\end{equation}

$=0.0065 A .$

In the work of \cite{IbenLivio1993} the different evolutionary outcomes for a close binary system are presented (Fig. 20) as regions in the plane of initial orbital separation $a_0$ (initial semimajor axis for a circular orbit) vs. $M_{1,0}$ (initial primary mass). In their calculations, for the primary to have a CO core the orbital separation range for our primary stellar mass range is between 1 (for the lower mass primaries) and 0.8 (higher mass) in log scale. So the relevant orbital separation range is $\sim 0.9$ in a log scale. We note though that when eccentricity is considered, even systems with an initial orbital separation of $a_0 \sim 10 \AU$
can go through the suggested evolutionary routes when the primary suffers a RLOF after developing a CO core.
As well, we can consider an earlier RLOF when the primary has a He core, as later the He core evolves to form a CO WD (similar to \citealt{TaurisSennels2000}). Over all we take the orbital separation relevant for the considered evolution to be 1 dex.
We take the initial orbital separation of the binary population of massive stars to span a range of 5 orders of magnitude
(from $a_{\rm min} \sim 10 R_\odot$ to $a_{\rm max} \sim 5000 \AU$) with an equal probability in the logarithmic of the orbital separation.
Accordingly, the probability of a binary system to be in the desired orbital separation is $f_s=1/5=0.2$.
This is a crude estimate as we took an order of magnitude value for the relevant orbital separation range and 
did not examine the detailed pre-common envelope evolution and its dependence on the initial separation, eccentricity, and the masses of the two stars.

The fraction of the systems studied here to the number of progenitors of CCSNe for $\eta=0.9$ is
\begin{equation}
\frac{N_{\rm WD-NS}}{N_{\rm CCSN}} \simeq  \frac { N_{b}}{N_{\rm CCSN}} f_{b} f_s \simeq 0.05.
\label{eq:frac1}
\end{equation}
Allowing for more mass loss (e.g., $\eta<0.9$) during the mass transfer process will reduce the number of systems as well. 
 Also, if the relevant orbital separation range is smaller, e.g., only 0.5 dex (say from 0.5\AU  to 1.5\AU), then the number of systems is half of that given above. 
To give a more conservative lower limit to our estimated number of systems we repeat the above calculation for $\eta=0.8$ and find 
\begin{equation}
\frac{N_{\rm WD-NS}}{N_{\rm CCSN}} \simeq  \frac { N_{b}}{N_{\rm CCSN}} f_{b} f_s \simeq 0.044.
\label{eq:frac2}
\end{equation}
Over all we take the fraction of the considered systems out of all CCSNe to be in the range of $3-5 \%$.
Using the CCSNe rate in our Galaxy from \cite{Cappellaroetal1997}, we estimate the Galactic birthrate of WD-NS reverse evolution systems to be  BR$_{\rm G}$(WD-NS) $ \simeq 5 \times 10^{-4} \yr^{-1}$.

We next compare our Galactic birthrate to previously derived Galactic birthrates.
\cite{PortegiesZwartVerbunt1996} considered systems where the initial mass of the primary star is $>8 M_\odot$ and its WD remnant has a mass in the range $1.1 M_\odot \la M_{\rm WD} \la 1.4 M_\odot$. They found in their population synthesis the Galactic birthrate of eccentric WD-NS binaries to be $\simeq 4.4 \times 10^{-5} \yr^{-1}$.
For comparison, we find the Galactic birthrate for our systems in the corresponding range (of $8M_\odot\lesssim M_{1,0}\lesssim 8.5M_\odot $) to be $\simeq 9\times 10^{-5} \yr^{-1}$.
Our estimate is twice that of \cite{PortegiesZwartVerbunt1996} mainly because we consider other evolutionary routes that leave no WD-NS binary systems. 
\cite{Kalogeraetal2005} estimated the Galactic birth rate of eccentric WD-NS binaries and find it to be $\simeq 0.3-1 \times 10^{-4} \yr^{-1}$. This is lower than our rate, but comparison is hard as they don't specify their parameter space. 
\cite{Daviesetal2002} estimated the Galactic birthrate of WD-NS, taking also into consideration cases where the system is expected to merge (as expected in J1141-6545 like systems) and found that the Galactic birthrate has a wide range of 
$\simeq 2\times 10^{-6}\yr^{-1}-4\times 10^{-4}\yr^{-1}$ and even up to $10^{-3}\yr^{-1}$ at high CE ejection efficiency. 
\cite{Nelemansetal2001} give an estimated Galactic birthrate of WD-NS binaries of $\simeq 2.4\times 10^{-4} \yr^{-1}$ and also estimate an expected Galactic merger rate from such systems to be $\simeq 1.4\times 10^{-4} \yr^{-1}$. We note that they do not differ between classic NS-WD evolution (where the NS is "born" first) and the reverse evolution. \cite{TaurisSennels2000} show that the WD-NS reverse evolution systems occur 10 times more than NS-WD evolution, hence we can crudely estimate the WD-NS Galactic birthrate of
\cite{Nelemansetal2001} to be  $\simeq 2.2\times 10^{-4} \yr^{-1}$.
Over all our Galactic birthrate estimation is close to those found in previous population synthesis studies.

We conclude that the WD-NS reverse evolution, including routes where the end point is not a WD-NS binary system, is rare but not negligible when peculiar SN explosions and ILOT events are considered. These systems are expected to be born at a fraction of $3-5$ percent of all CCSNe and at a Galactic rate of $\simeq 5 \times 10^{-4} \yr ^{-1}$.

% ====================
\section{COMMON ENVELOPE EJECTION}
\label{sec:CE}
% ====================

To examine whether the envelope of the secondary star is ejected by the spiraling-in WD we compare the binding energy of the envelope residing above radius $a$
\begin{equation}
E_{\rm bind} =- \int_{M(a)}^{M(R_\ast)}  \left( e_{\rm G} + e_{\rm int} \right) dm
\label{eq:bind_e}
\end{equation}
where $R_\ast$ is the stellar radius, with the energy liberated by the spiralling-in WD
\begin{equation}
E_{\rm orb} =\frac{1}{2} \frac {GM({\rm r}) M_{\rm WD}}{a}.
\label{eq:eorb}
\end{equation}
Here $e_G$ and $e_{\rm int}=e_{\rm th}+e_{\rm rad}$ are the gravitational and internal energy per unit mass, respectively, where the internal energy is composed of thermal $e_{\rm th}$ and radiation $e_{\rm rad}$ energies.
Lower case `e' will stand for energy per unit mass while upper case `E' will mark the total corresponding energy type of the envelope outside radius $r$.
In the last equation we take the WD to start from a large radius outside the giant photosphere. Substituting the thermal and radiation energy per unit mass into the internal energy in equation (\ref{eq:bind_e}) we get:
\begin{equation}
E_{\rm bind} =-\int_{M(a)}^{M(R_\ast)} \left( e_{\rm G} + e_{\rm th} + e_{\rm rad} \right) dm
.
\label{eq:ebind}
\end{equation}

\begin{figure}[ht]
\centering
\includegraphics[width=79mm]{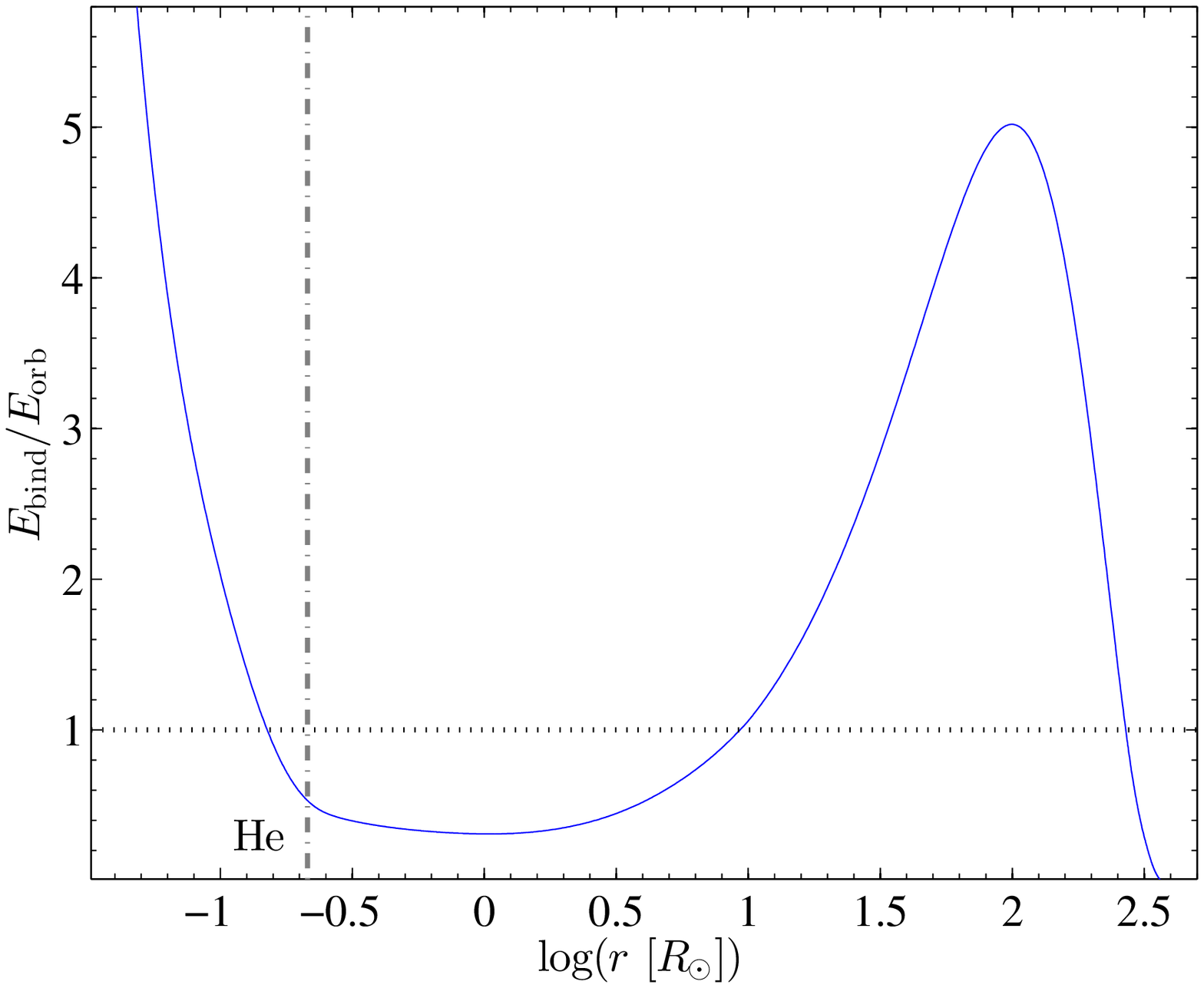}
\hspace*{1mm}
\includegraphics[width=80.5mm]{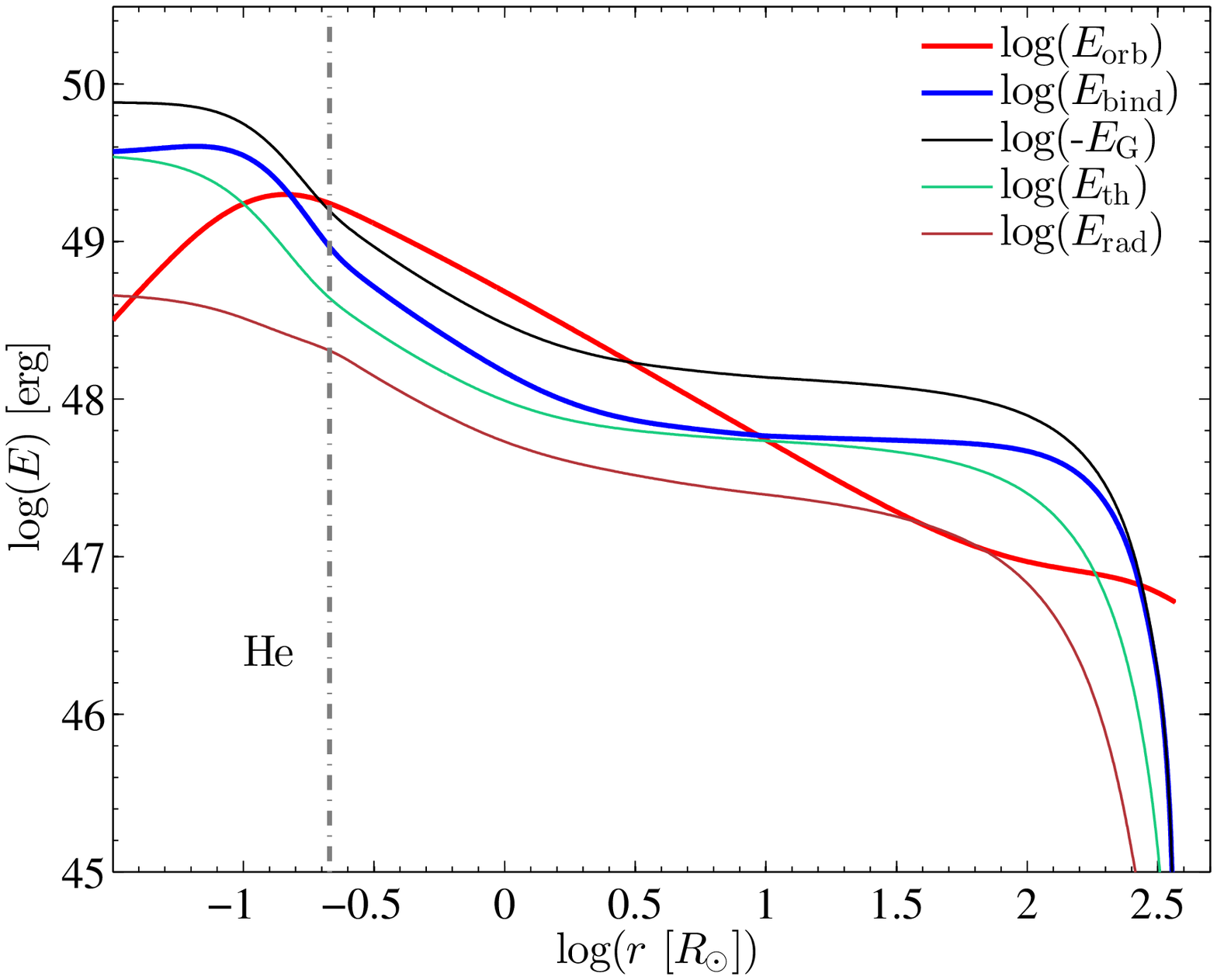}
\centering
\includegraphics[width=80.5mm]{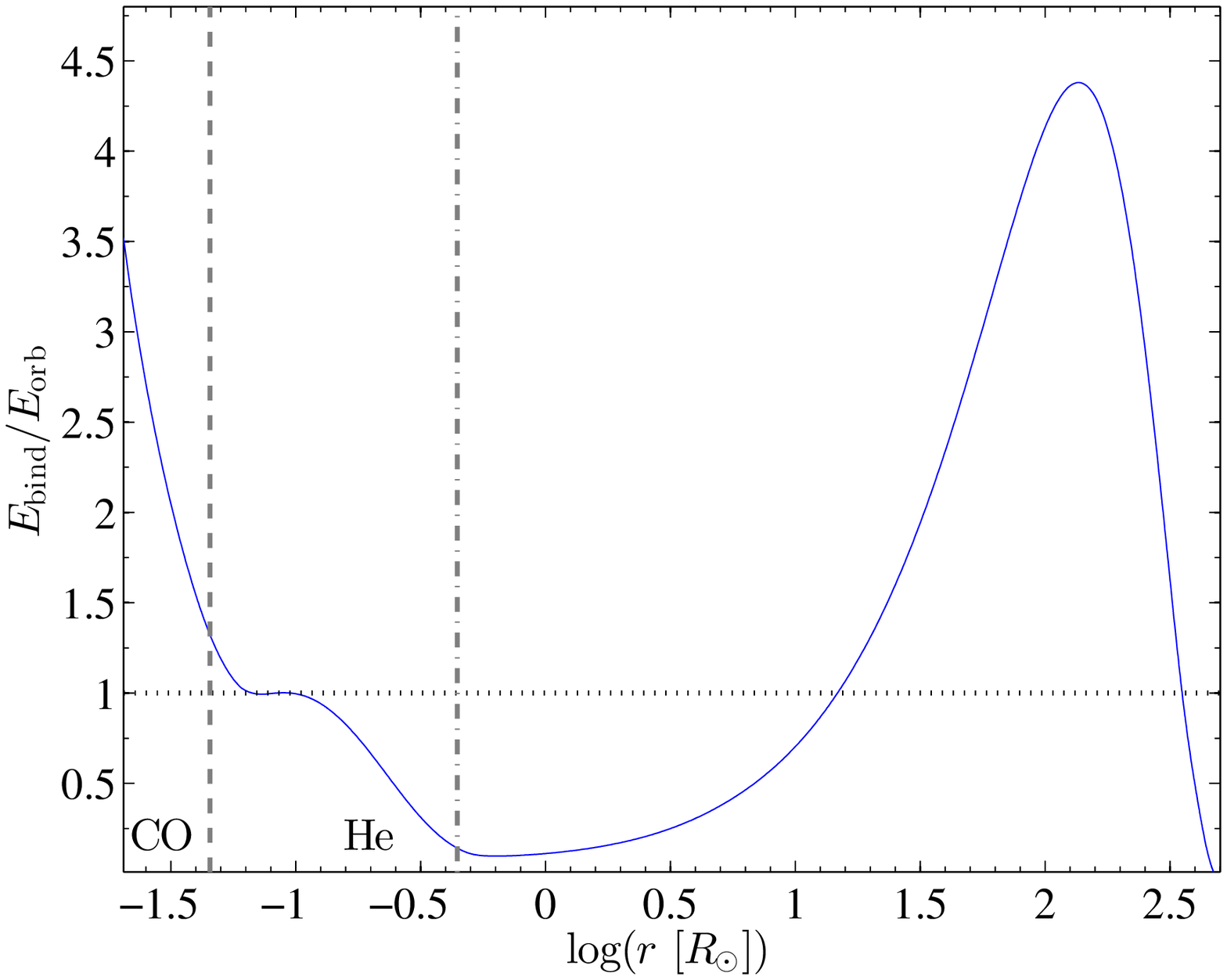}
\hspace*{1mm}
\includegraphics[width=80.5mm]{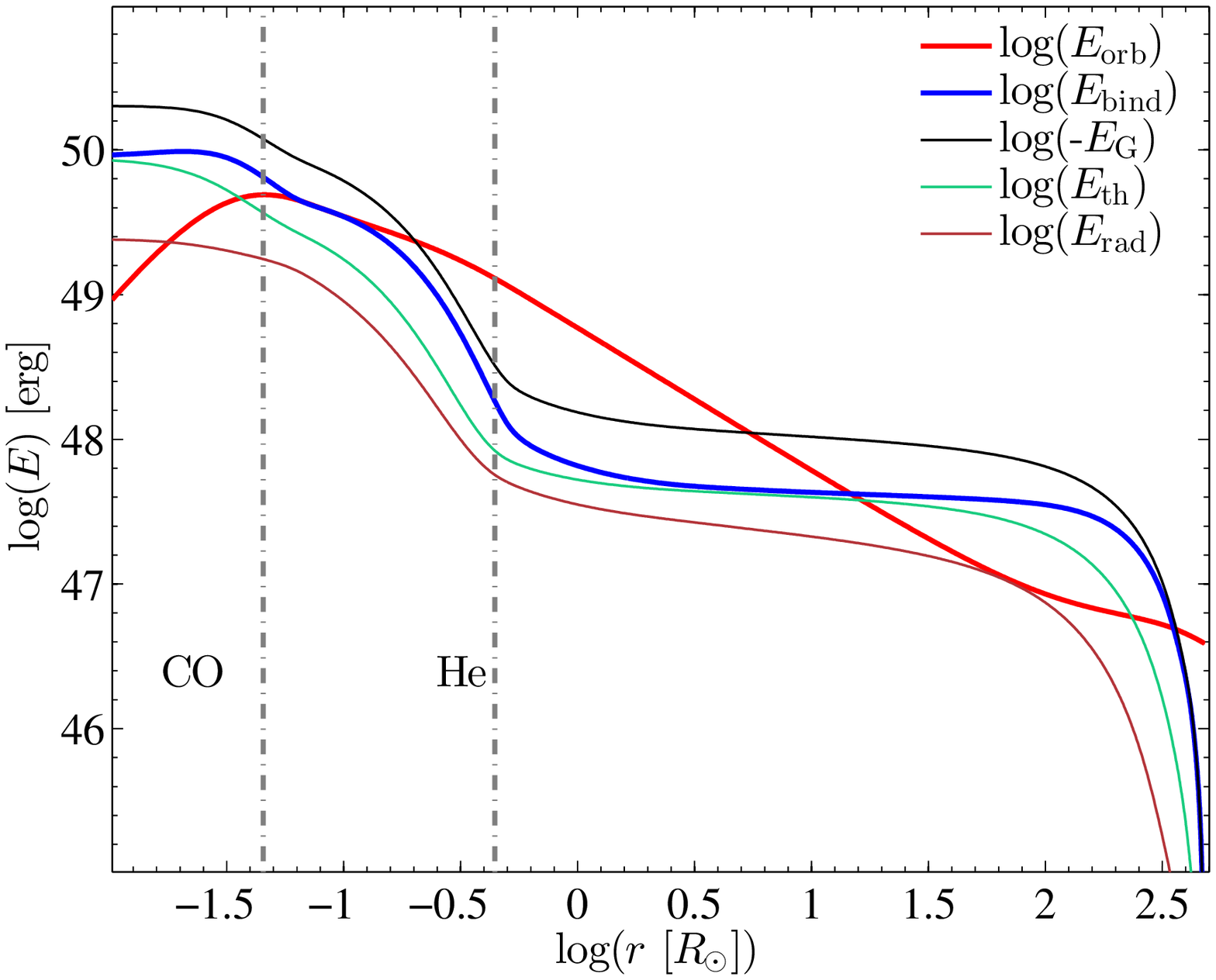}
\caption{Left panels: The ratio of the binding energy of the envelope residing above radius  r  to the released orbital energy when the orbital separation is $a={\rm r}$. The WD mass is $M_{\rm WD}=1 M_\odot$, and the giant is of $10M_\odot$. Right panels: The different energies calculated for the spiraling-in phase:
$E_{\rm orb}$  is the orbital energy released by the spiraling-in companion (equation \ref{eq:eorb}) when the orbital separation inside the envelope is $a$.
$E_{\rm bind}({\rm r})=-[E_{\rm G}({\rm r}) +E_{\rm int}({\rm r})]$ is the binding energy (equation \ref{eq:bind_e}), where $E_{\rm int}({\rm r})=E_{\rm th}({\rm r}) + E_{r\rm ad}({\rm r})$ is the internal energy, $E_{\rm th}({\rm r})$ is the thermal energy,  and $E_{\rm rad}({\rm r})$ is the radiation energy; all for the envelope residing above radius ${\rm r}$. These are presented for when the giant has developed a massive He core (upper panels) and for the case of a massive CO core (lower panels).  The horizontal dashed line in the left panels is to guide the eye. }
\label{fig:E_b10}
\end{figure}

\begin{figure}[ht]
\centering
\includegraphics[width=80mm]{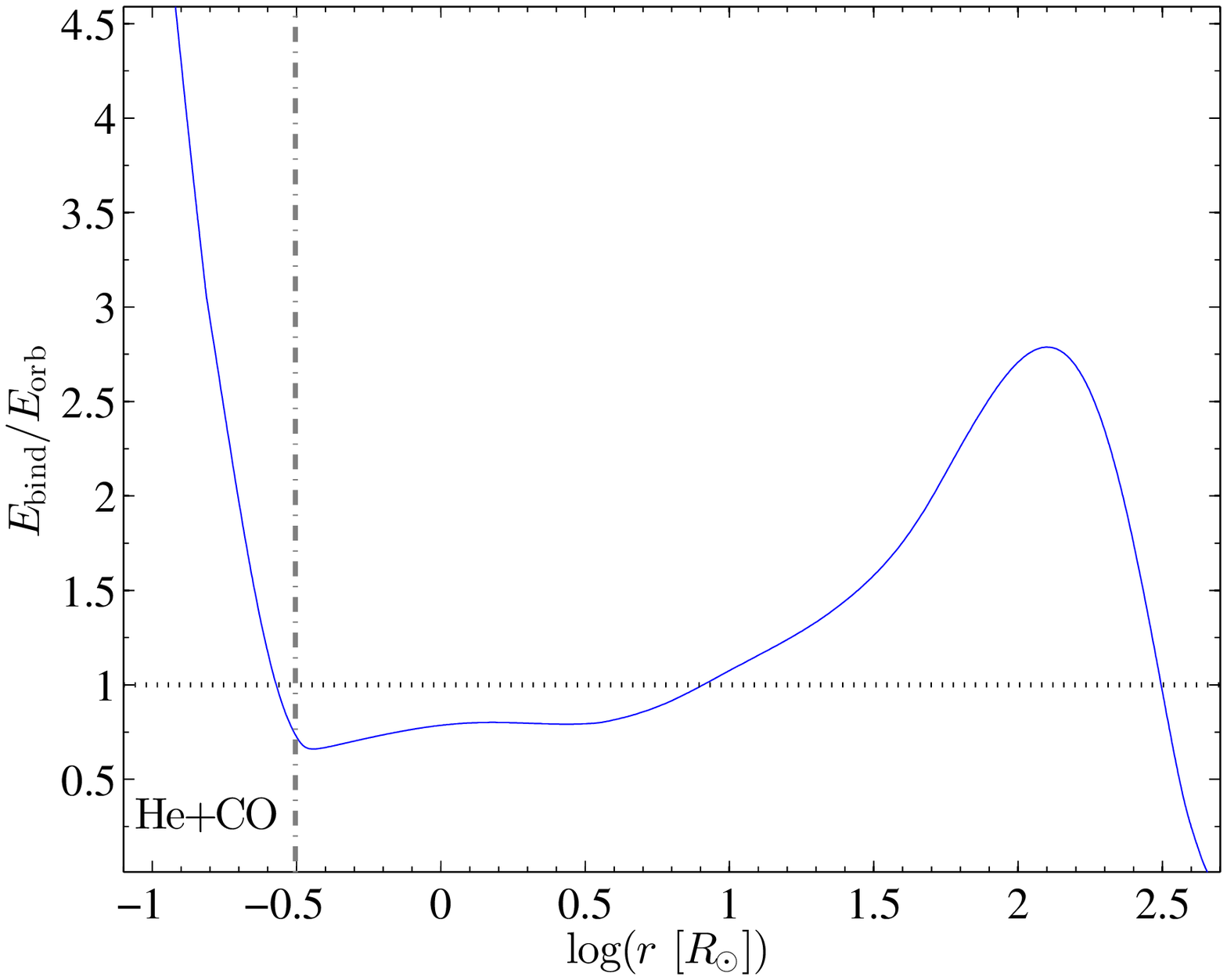}
\hspace*{1mm}
\includegraphics[width=80mm]{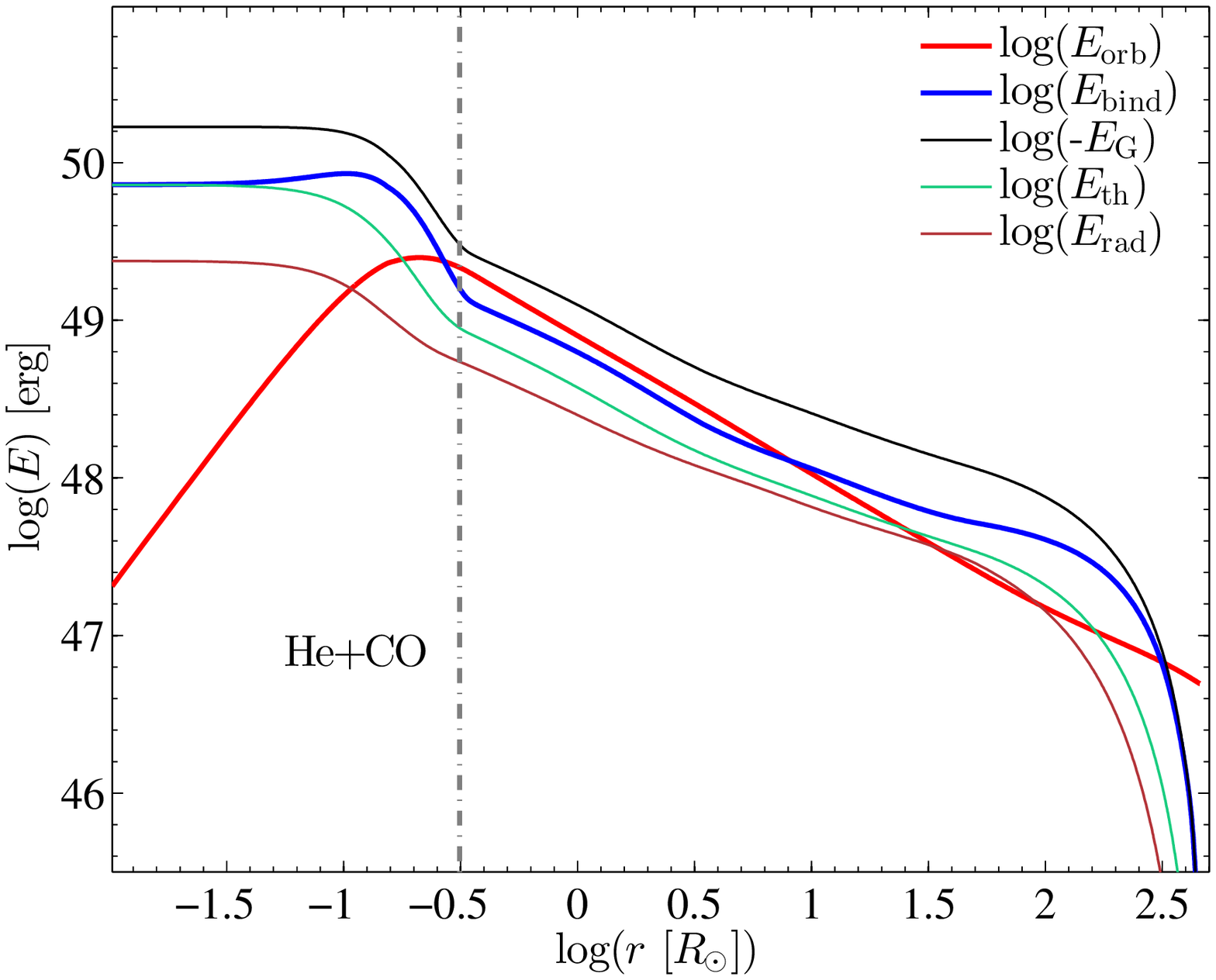}
\includegraphics[width=80mm]{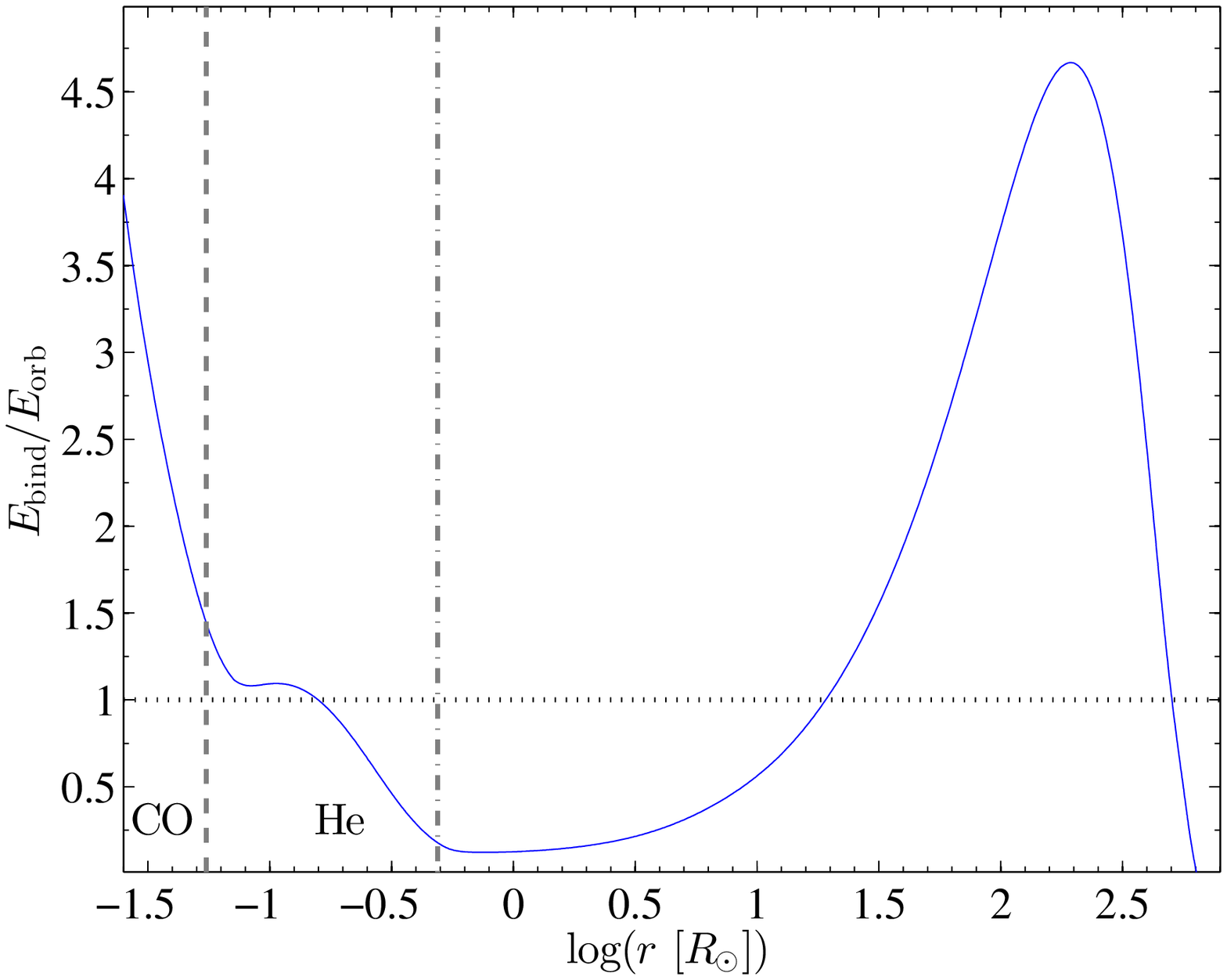}
\hspace*{1mm}
\includegraphics[width=80mm]{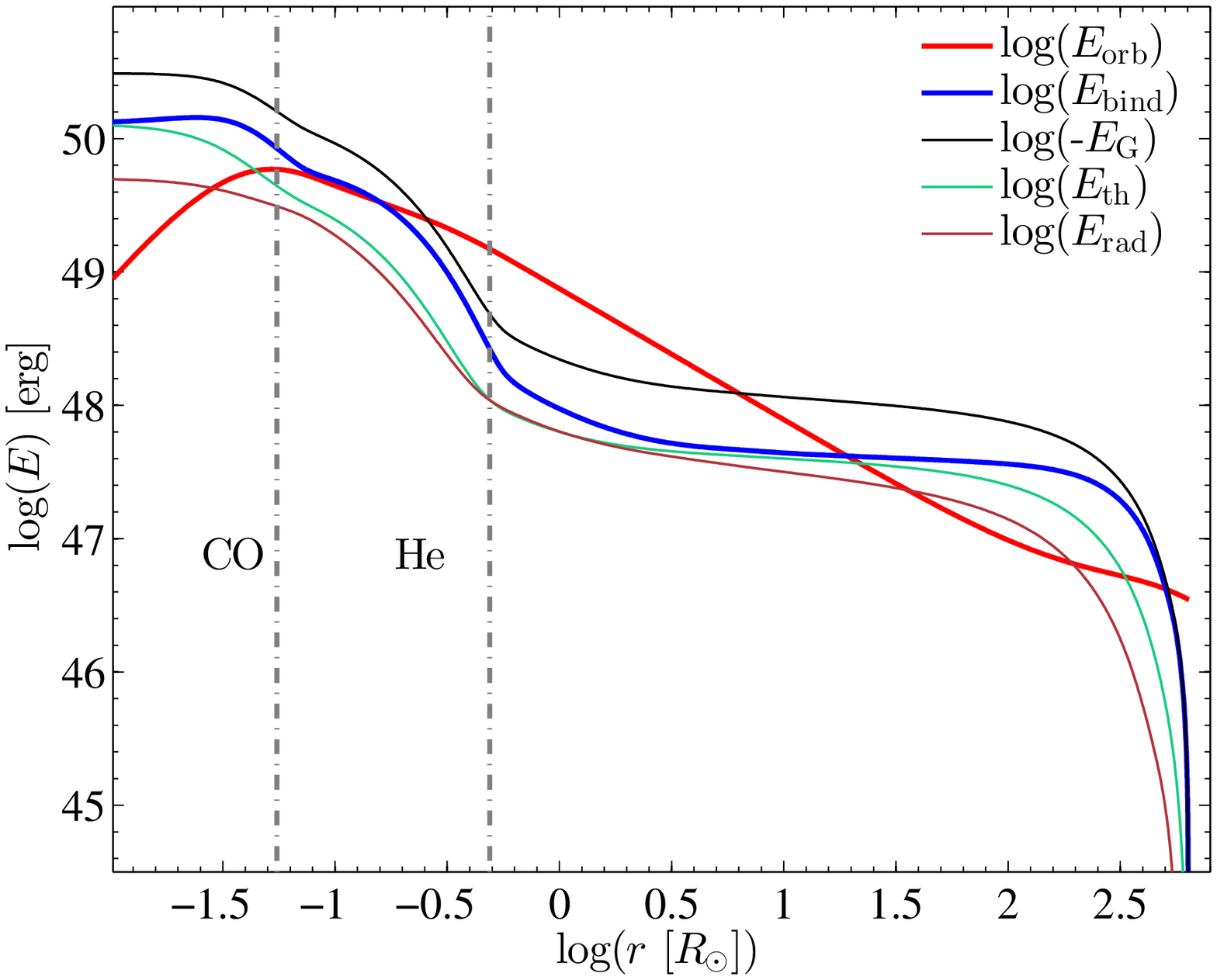}
\caption{Like Fig. \ref{fig:E_b10} but for the $12M_\odot$ model when the star has a massive HeCO core (upper panels) and a massive CO core (lower panels). }
\label{fig:E_b_12}
\end{figure}

The values of these energies for the $10M_\odot$ model with He and CO cores are given in Figs. \ref{fig:E_b10},
and for the $12M_\odot$ model with a HeCO and CO cores in Fig. \ref{fig:E_b_12}.
In calculating the orbital energy we take a WD of mass $M_{\rm WD}=1M_\odot$.
We note that this approach is somewhat different
from the usual practice of taking the binding energy of the entire envelope, from the core radius to the surface, as discussed by \cite{DewiTauris2000} (for a recent study of CE evolution and the standard energy formalism see \citealt{Ivanovaetal2013b}).
We examine here the radius at which the WD expels the envelope outside its location. The difference from the commonly used prescription for envelope ejection is not large when the WD is very close to the core, as are the cases discussed here.

As shown in the work of \cite{TaurisDewi2001} the outcome of a CE in a binary system depends on the exact location of the bifurcation point$-$the layer separating the ejected envelope from the core region$-$which defines the core mass of the donor star ($M_2$ in our case). They  compared various methods in defining the bifurcation point and found for their studied CE donor stars, of masses $4M_\odot,\;7M_\odot,\;10M_\odot$ and $20M_\odot$, that although it is straightforward to define this point for a donor star at the AGB phase, for an RGB star the different methods result in different core masses.
Though this is an important point, we here do not treat the CE in the standard manner and thus do not take the core mass into consideration in estimating the final orbital separation of the system.
Our treatment of the binding energy down to the location of the secondary rather than the binding energy of the envelope,
whose inner boundary is not well defined \citep{TaurisDewi2001}, actually avoids this problem.
We only consider the core mass and radius in our further estimation of the merger energy (eq. 9), where we define the core of the giant secondary
(the "bifurcation point") as the inner part of the star which is hydrogen poor.

% FFFFFFFFFFFFFFFFFFFFFFFFFFFFFFFFFFFFFFFFFFFFFFFFFF
% FFFFFFFFFFFFFFFFFFFFFFFFFFFFFFFFFFFFFFFFFFFFFFFFFF
% FFFFFFFFFFFFFFFFFFFFFFFFFFFFFFFFFFFFFFFFFFFFFFFFFF
% FFFFFFFFFFFFFFFFFFFFFFFFFFFFFFFFFFFFFFFFFFFFFFFFFF

It is evident from Figs. \ref{fig:E_b10} and \ref{fig:E_b_12} that in the cases studied here there is a zone inside the giant where $E_{\rm bind}/E_{\rm orb} \la 1$.
When the WD reaches this zone it might in principle eject the envelope outside its location, but not by a large margin.
On the one hand there is extra energy available for envelope ejection, e.g., radiation that powers the stellar wind.
On the other hand the efficiency of channelling the released orbital energy to the envelope ejection is $\alpha_{\rm CE} <1$.
Also, some portion of the envelope might leave the star with energies much above their binding energy \citep{KashiSoker2011},
leaving behind bound circumbinary disk that can lead to a merger \citep{KashiSoker2011, Soker2013a}.
These processes are poorly determined, and for that we consider both the possibility of complete and partial envelope ejection,
as depicted in Figs. \ref{fig:mech1} and \ref{fig:mech2}. These outcomes will be further discussed in section \ref{sec:postCE}.

In case a merger does occur, an interesting quantity for the discussion to follow in section \ref{sec:postCE} is the difference between the
orbital energy released at WD-core merger and the binding energy of the entire envelope.
\begin{equation}
E_{\rm m} = E_{\rm orb}(R_{\rm core})-E_{\rm bind}(R_{\rm core})
\label{eq:Emerger1}
\end{equation}
A fraction of this energy will be radiated in an Intermediate Luminosity Optical Transient (ILOT) event.
This quantity for the four models considered here is given in Table \ref{tab:parameters}.

%%TTTTTTTTTTTTTTTTTTTTTTTTTTTTTTTTTTTTTTTTTTTTTTTTTTTTTTTTTTTTTTTTTTTTTTTTTTTTTTTTTT
\begin{deluxetable}{lcccc}
%%\rotate
\centering
\tablecolumns{5}
\tablewidth{0pc}
\tablecaption{The parameters of each system at the CE phase}
\tablehead{
    \colhead{Model} &
    \colhead{$10M_\odot$} &
    \colhead{$10M_\odot$} &
    \colhead{$12M_\odot$} &
    \colhead{$12M_\odot$} \\
\colhead{ } & \colhead{He core} & \colhead{CO core} & \colhead{HeCO core} & \colhead{CO core}
}
\startdata
   $M_{\rm core}\;[M_\odot]$ & 1.965 & 3.02 & 3.57 & 3.83
\\ $\mu\;[M_\odot]$ &  0.66 &  0.75 & 0.78 &0.79
\\ $R_{\rm core}\;[R_\odot]$ &  0.21 &0.44 & 0.31 & 0.49
\\ $t_i \;[10^7{\rm yr}]$ &8.46 & 8.67& 6.08 &6.14
\\ $\delta t\;[\rm yr]$ &$2.2\times 10^6$ &$3.7\times 10^4$ &$6.3 \times 10^5$  &$1.5\times 10^4$
\\ $E_{\rm m}\;[10^{48}{\rm erg}]$ & 8.18 & 11.05 & 5.72 & 12.45
\\ $I_{\rm core}\;[M_\odot R_\odot^2]$ &0.018 & 0.043& 0.069 &0.064
\\ $\xi$ & 0.60 & 0.29 & 0.905 & 0.34
\\ $a_{s}$ &  0.945 & 0.885 & 0.823 &0.838
\\ $a_{D}$ & 0.287& 0.415 & 0.515 & 0.492
\enddata
\footnotesize
\flushleft
{$\mu$ is the reduced mass of the core and a WD of $M_{\rm WD}=1M_\odot$;
$t_i$ is the age of the system at the CE phase, measured from the time at which the post-accretion secondary had relaxed on the MS;
$\delta t$ is the remaining lifetime of the giant to explosion had there been no CE phase;
$E_{\rm m}$ is the orbital energy released in the WD-core merger process minus the envelope binding energy as given in equation (\ref{eq:Emerger1}) and for $M_{\rm WD}=1M_\odot$;
$I_{\rm core}$ is the moment of inertia of the core, as defined in the Appendix;
$\xi$ is defined in equation (\ref{eq:xi1}) of the Appendix;
$a_s$ is the distance at which the WD brings the core to synchronization from an initial post-CE orbital separation of $a_f=1 R_\odot$, as defined in the Appendix;
$a_D$ is the distance at which the system is Darwin unstable, as defined in the Appendix.
}
\\[1.5ex]
\label{tab:parameters}
\end{deluxetable}
%%TTTTTTTTTTTTTTTTTTTTTTTTTTTTTTTTTTTTTTTTTTTTTTTTTTTTTTTTTTTTTTTTTTTTTTTTTTTTTTTTTT

In cases we term complete envelope ejection the envelope outside radius $a_f$ is ejected by the released orbital energy,
as this is our approach to CE ejection.
Some H-rich gas is left between the core and the WD orbit, that might cause spiraling-in by tidal interaction.
It is hard to estimate this mass as the envelope adjusts itself during the spiraling-in process and inner layers
expand and their binding energy will substantially decrease.
As shown in Fig \ref{fig:density} there is not much mass between the core and $r\simeq 1R_\odot$ to begin with,
and we assume that in some cases the entire envelope is then ejected, not just the mass above $a_f$,
hence our terminology of complete envelope ejection.
The core readjustment might lead to rapid mass transfer and merger \citep{Ivanova2011}.
Therefore, in many cases of complete envelope ejection later WD-core merger will occur accompanied by mass transfer from the core to the WD.
The outcomes of mass transfer and merger will be discussed in section \ref{sec:postCE}.

In cases of complete envelope ejection the orbital separation will be $a_f \simeq 0.3-1 R_\odot$, as evident from Figs. \ref{fig:E_b10} and \ref{fig:E_b_12}.
At this stage further spiraling-in might occur due to two processes.
First, bound envelope material might fall back and interact with the binary system, e.g., via a circumbinary disk \citep{KashiSoker2011}.
 This process might take weeks to years, as the dynamical time for the fall back material to reach the center.
Second, if the rotation (spin) of the core is not synchronized with the orbital motion, tidal forces will act to bring synchronization.
There are very large uncertainties in determining the synchronization time, in particular as the exact convective structure of the exposed
core must be known.
We found (see Appendix) that even if tidal interaction is strong and acts on a short time scale, synchronization is achieved after the orbital separation has decreased by a very small fraction, and in most cases the system is then stable to the Darwin instability \citep{Darwin1879}.  
We therefore conclude that tidal interaction has negligible effects on the post-CE system (unless the core re-expands).

The conclusion from the above discussion is that if the WD manages to eject the entire envelope,
the WD-core binary system will survive at least until further core evolution, or, if there is a fall back gas, a circumbinary disk might cause a merger.
This further evolution is discussed next.

% ====================
\section{POSSIBLE OBSERVATIONAL SIGNATURES}
\label{sec:postCE}
% ====================

The final outcomes of these systems will be one or two SN explosions, and in some cases ILOTs.
There are large uncertainties regarding the occurrence and properties of the explosions.
On a more solid ground is the occurrence of ILOTs, sometimes preceding explosions and termed pre-explosion outburst (PEO), in
many of the routes. We therefore start by estimating the ILOT properties.
%\footnote[1]{see also \href{http://physics.technion.ac.il/~ILOT/}{as}}.

% ============
\subsection{Intermediate luminosity optical transient (ILOT)}
\label{sec:ILOT}
% ============

In cases when the WD merges with the core the amount of gravitational energy released at the final spiraling-in phase is larger than the
binding energy of the entire envelope by $E_{\rm m} \simeq 10^{49} \erg$, as given in the sixth row of Table \ref{tab:parameters}.
This energy will be channelled mainly to the kinetic energy of the ejected envelope, and some fraction of it to radiation.
If the ejected gas collides with gas ejected earlier in the CE process, then more energy will be radiated.
The duration of the outburst will be determined by the diffusion time of photons from the ejected gas, as in SNe.
In this case the outburst lasts for several weeks. If it is further powered by collision of ejecta, it might last for
few months and have a complicated light curve with more than one peak.
Over all an event with radiated energy of $\sim 10^{48} - 10^{49} \erg$ and lasting for $\sim 1-10$~weeks might take place.
This is defined as an ILOT event.
For the systems studied here such an ILOT will be followed by an explosion, a PEO.
The PEO of SN~2010mc observed by \cite{Ofeketal2013} has properties similar to the ILOTs proposed
here\footnote[1]{see also \url{http://physics.technion.ac.il/~ILOT/}}.

The routes that might have an ILOT/PEO can be identified in Figs. \ref{fig:mech1} and \ref{fig:mech2}.
For the CO WD these are CO-1,~2,~5,~6 and for the ONeMg WD these are ONe-1,~2,~3,~6,~7,~8.

It is beyond the scope of the present paper to examine the exact properties of the ILOT/PEO that will result from a
WD-core merger process, as it requires to follow the ejection mass history and to include radiative transfer.
Also, the time delay between the ILOT and explosion, which can be days to many years, requires deeper calculations.
It would be better to examine specific type II SNe that have a PEO, and try to constrain the energy and mass involved in the PEO;
some of these PEOs might be accounted for by scenarios discussed here.

% ============
\subsection{Thermonuclear explosion}
\label{sec:thermo}
% ============

The ignition of thermonuclear explosions will not be studied here.
We rather limit ourselves to speculate on the routes where thermonuclear explosion might take place.
For the CO WD these are routes CO-1,~3,~5 and for the ONeMg WD these are routes ONe-2,~7.
Core explosion could occur if there is a CO-rich core. When the core of the giant companion is He-rich the core will not go through a thermonuclear explosion.

In routes CO-1 and ONe-2 thermonuclear explosion occurs inside the envelope.
In CO-1 the WD will explode, and in ONe-2 explosion of CO accreted onto the ONeMG WD might explode if the core is CO-rich; ignition
itself might be by accreted helium.
As the ignition is by accreted core material, CO or more likely He, and since it is possible that part of the core material will
go through a thermonuclear outburst as well, we indicate in Figs. \ref{fig:mech1} and \ref{fig:mech2} `Explosion of WD and/or core' for these channels.
In these cases there will be wide hydrogen lines and the SN will be classified as a Type II.
As there is a massive circumstellar gas, most likely it will be a Type IIn SN.
However, the driving engine is not a CCSN, and these will be peculiar Type II SN, e.g., having a large mass of nickel and other synthesized elements.

In routes CO-5 and ONe-7 the WD-interaction with the core might ignite the core (and/or the WD in CO-5),
as in the violent merger ignition \citep{Pakmoretal2011, Pakmoretal2012}. Basically, when a WD accretes violently He or CO, thermonuclear explosion might occur on the surface of the WD that ignites the entire WD. As the WD is very close to the core or even inside it, some of the core material might also be ignited. Furthermore, it is possible that the core material itself will be ignited from the violent merger of the core and the WD.
Even if the initial core was He-rich, the merger itself might take place after a massive CO central core develops; this will be studied in detail
in a future study.
These routes might occur long after the H-rich envelope has dispersed and there will be no narrow absorption lines.
However, as the WD enters the CO core, or a somewhat more evolved core, it ejects some helium. The helium will be ejected at high speeds
(escape velocity from the core).
Such an explosion, if occurs, will be found to have massive ejecta and helium lines.
It will be classified as a peculiar Type Ib SN as it will have massive Ni ejecta.
In route CO-3 the WD accretes mass from the core, a process that might lead to a SN Ia
where the remnant is a bright exposed core, e.g., a WR star.
If this route occurs long after the CE phase, the WD explosion will be classified as a typical SN Ia,
unless there is a fast wind ($\sim 10^3 \km \s^{-1}$) from the exposed core, now a WR star,
that will be observed as weak medium-velocity absorption lines.

Routes CO-3,~5 and ONe-7 might also occur shortly after the CE phase, when the ejected CE is not far from the star.
Narrow absorption lines of hydrogen and other element will be observed.
Route CO-5 deserves a much deeper study.
If all the He burns, then it will be classified as SN Type Ia with massive ejecta, $\sim 3-5 M_\odot$, and large amount of nickel.
If the He survives the explosion, then this will be classified as a peculiar Type Ib SN.
In route ONe-7 the ONeMg WD is likely to survive the thermonuclear explosion of the core, or else collapse to a NS.
There is of course the question whether such a
thermonuclear explosion of the core will take place when the ONeMg WD spirals inside.

% ============
\subsection{Core collapse SNe (CCSNe)}
\label{sec:CCSNe}
% ============

We now speculate on the routes where CCSNe might take place, though this also will not be studied here in detail.

In cases where the massive core of the giant secondary star gravitationally collapses, triggered by electron capture or a massive Fe core,
and the hydrogen envelope had already been
completely ejected during the CE phase, the explosion will be classified as a CCSN Type Ib or Ic (Type Ibc).
For a CO WD these are routes CO-3,4,6 and for the ONeMg WD these are routes ONe-4,5,6,8.
These routes might also occur when the ejected CE is not far
from the star and narrow absorption lines of hydrogen and other elements will be observed.
These will be classified as Type IIn SNe, or Ibc with narrow H lines.

In case of only partial envelope ejection, routes CO-2 and ONe-1,3 the collapse occurs inside the H-rich envelope.
In these cases wide hydrogen lines will be observed, and the explosion will be classified as a Type II SN.

Another possible route is the one of accretion induced collapse (AIC; for a recent study of AIC see \citealt{Taurisetal2013}).
It is likely to happen for an ONeMg WD.
This can take place when the in-spiral of the WD ends outside the core and RLOF occurs, as in route ONe-4.
Alternatively it can occur in the case of WD-core merger, where the AIC happens inside the core, as in routes ONe-1,6.
When the NS forms a huge amount of gravitational energy is liberated, and the explosion will be classified as Type Ibc if there is no
H-rich envelope, as in routes ONe-4,6.
However, the SN will be peculiar since the core around the collapsed WD did not finish its nuclear evolution, unlike regular CCSN.
If there is a H-rich envelope, route ONe-1, a peculiar Type II CCSN SN will occur.

We note that the final outcome of routes CO-4 (WD+NS), ONe-4 (NS+NS) and ONe-5 (WD+NS) are binary systems. 
Such reverse evolution that lead to these types of binary systems was discussed in the past in several studies 
(e.g.,  \citealt{TutukovYungelSon1993, PortegiesZwartVerbunt1996, vanKerkwijkKulkarni1999, PortegiesZwartYungelson1999, TaurisSennels2000, Brownetal2001, Nelemansetal2001, Daviesetal2002, Kimetal2003, Kalogeraetal2005, Churchetal2006, vanHaaftenetal2013}).

% ====================
\section{DISCUSSION AND SUMMARY}
\label{sec:Summary}
% ====================

We examined rare evolutionary routes where in a binary system a white dwarf (WD) is born before the neutron star (NS), or a massive core that might become a NS progenitor, in what we term \textit{WD-NS reverse evolution}.
This is made possible by a mass transfer from the initially more massive primary star to the secondary star.
The mass ranges of the two stars and the different possible outcomes are summarized in Figs. \ref{fig:mech1} and \ref{fig:mech2}.
The formation of a WD-NS system where the WD is born first was mentioned before with emphasis on the final outcome of a binary WD-NS system
(e.g.,  \citealt{TutukovYungelSon1993, PortegiesZwartVerbunt1996, vanKerkwijkKulkarni1999, PortegiesZwartYungelson1999, Brownetal2001, Nelemansetal2001, Daviesetal2002, Kimetal2003, Kalogeraetal2005, vanHaaftenetal2013}).
\cite{TaurisSennels2000} and \cite{Churchetal2006} give a detailed evolution of a WD-NS system to explain the binary pulsars PSR B2303+46 and 
PSR J1141-6545. They have some assumptions similar to ours, e.g., that merger might occur, although they do not emphasize this outcome.
Here we considered the WD-NS binary outcomes, as well as merger, the possibility of an ILOT event, and a  variety of peculiarities.

In the studied scenarios the more massive star evolves first to a WD, but the secondary star is in an orbital distance
that facilitates a Roche lobe overflow (RLOF) during the AGB phase of the primary star or at an earlier stage when the primary has a He core.
In the later case the post-transfer primary shrinks, mass transfer ceases, and the He core evolves to form a CO (or ONeMg) WD (similar to \citealt{TaurisSennels2000}).
In the majority of cases the WD descendant of the primary star is a carbon-oxygen (CO) WD, but when the initial mass of the primary is
in the range $M_{1,0}\approx 7-8.5M_\odot$ the remnant might be an ONeMg WD.
The post-accretion secondary becomes massive enough, $M_2 > 8.5 M_\odot$, to be considered a progenitor of a core collapse supernova (CCSN).

Routes where the primary star leaves a helium WD, because RLOF occurs very early in the evolution and the He core is not massive enough to form a CO WD, and the post-accretion secondary star becomes a CCSN also exist, but were not discussed in our study. We expect that the low mass helium WD will merge with the secondary core after the secondary becomes a giant and before complete envelope ejection, and will not lead to any peculiar explosion. However, an ILOT event preceding a CCSN by a long time will take place.

Using the powerful MESA tool for stellar evolution \citep{Paxton2011}, we calculated the evolution of primary stars and
post-accretion secondary stars (Figs. \ref{fig:7MS} - \ref{fig:HR_comp}).
The post-accretion secondary star starts its main sequence (MS) stage as a helium-enriched star.
Two cases of the evolution of the post-accretion secondary star are presented in Figs. \ref{fig:ev_10} and \ref{fig:ev_12}.

As the post-accretion secondary expands to become a giant the WD cannot bring the envelope to synchronization and a
common envelope (CE) phase begins, causing the WD to spiral-in.
We examined whether the envelope of the secondary star is ejected by the WD by comparing the binding energy
 of the envelope residing above the location of the WD with the energy liberated by the spiralling-in WD.
Our approach to CE ejection is not the common one  (e.g., \citealt{Ivanovaetal2013b} and \citealt{ToonenNelemans2013}, and references therein), as we did not examine the binding energy of the entire envelope, but rather the binding
energy of the envelope residing above the location of the WD (eq. \ref{eq:ebind}), and the orbital energy at that separation (eq. \ref{eq:eorb}).
These two energies and their ratio for the two post-accretion models studied here are given in Figs. \ref{fig:E_b10} - \ref{fig:E_b_12}.
We find that envelope ejection can be marginally achieved when the orbital separation is $a \sim 0.3 -1 R_\odot$. At these short orbital separations our approach to CE ejection gives similar results to the common prescription of CE ejection.
If ejection does not occur, the WD and the core will merge while the giant still holds a part of its envelope.

To bring the core to synchronization the WD needs to further spiral-in only a small distance.
Furthermore, the synchronized system is practically stable against Darwin instability (eq. \ref{eq:Darwin1}, Appendix).
However, further core evolution and fall back gas can lead the system to merge.
In this case merger occurs with no envelope, but possibly with an optically-thick wind (the ejected envelope).
If merger occurs much later, then there will be no hydrogen in the vicinity.

The merger itself releases gravitational energy much larger than the binding energy of the envelope, as given in Table \ref{tab:parameters}.
Even if only a small fraction of this energy is radiated, the process will form a bright event lasting days to months.
In addition, the ejected gas from the merger process will collide with previously ejected envelope, and kinetic energy will be channelled to radiation
(see section \ref{sec:postCE}).
Such an event might be classified as an Intermediate Luminosity Optical Transient (ILOT; Red Nova; Red Transient).
This ILOT will be followed by a supernova (SN) explosion, in some cases more than one.
All the evolutionary routes end with one or two explosions, some of which will be classified as peculiar types in their group.
In most cases the thermonuclear explosion will occur weeks to years after an ILOT event, while CCSNe are expected to occur much later, $\sim 10^4-10^6 \yr$ after the core has finished its evolution (see Table \ref{tab:parameters}). Only the case with an AIC of an ONeMg WD a CCSNe type explosion might occur shortly after an ILOT event. If an explosion occurs only few days after an ILOT event, it might be that the two events will be observed as one explosion.

Our main results are that the type of binary systems studied here can lead to a rich variety of peculiar explosions, most of which will be preceded by
an ILOT event, and be accompanied by narrow lines of hydrogen from the previously ejected hydrogen rich envelope. These add to the rich variety of more traditional binary evolutionary routes of exploding stars (see review by \citealt{Langer2012}).
We estimate that the systems studied here occur at a rate of  $\sim 3-5$ percent of that of CCSNe, and at a Galactic birthrate of $ \simeq 5 \times 10^{-4} \yr^{-1}$ (see section \ref{sec:BR}).
As a diversity of CCSNe are being discovered in recent years (e.g., \citealt{Arcavi2012}),
it could be that some rare types can be accounted for by one of the routes studied here.
This is a subject of a future study.

The main observational peculiarities expected from these SNe are as follows.
\begin{enumerate}
\item As noted also by the papers cited above, if the binary system is still a part of a stellar cluster, an explosion of a massive star will be observed in a stellar cluster whose turn-over mass is $<8M_\odot$.
\item If WD-core merger occurs after partial envelope ejection, it could result in a thermonuclear explosion leading to a peculiar
type II SNe with massive ejecta of Ni and other synthesized elements. This is seen in routes CO-1 and ONe-2
in Figs. \ref{fig:mech1} and \ref{fig:mech2}, respectively.
\item If merger of an ONeMg WD with the core occurs after partial envelope ejection, the accretion of the core material onto
the ONeMg WD could result in accretion induced collapse (AIC) inside the core and a peculiar CCSN.
This is peculiar in the sense that the collapse occurs before the rest of the core had evolved as in regular CCSNe (ONe-1).
In cases where this happens after complete ejection of the H-rich envelope, this route will end in a peculiar CCSN Type Ibc (route ONe-6).
\item If merger of an ONeMg WD with the core occurs after entire envelope ejection a thermonuclear explosion of the core alone
will leave behind an ONeMg WD or a NS (if the WD goes through AIC). This might be classified as a peculiar SN type Ia or Ibc
(route ONe-7).
\item In route CO-3 there might be two peculiarities. First, there will be a massive H-rich circumstellar medium (CSM)
around the exploding star that is likely to be classified as Type Ia. Second, a very luminous remnant with
$L \simeq 5 \times 10^4 L_\odot - 8 \times 10^4 L_\odot$ will be left behind, until it  experiences a CCSN event on its own.
\end{enumerate}

We conclude that WD-NS reverse evolution, including routes where the end point is not a WD-NS binary system, is rare but not negligible when peculiar SN explosions and ILOT are considered, and are expected to occur at a rate of $\simeq 3-5$ percent of that of CCSNe.

\textit{Acknowledgements.} We thank Amit Kashi, Hagai Perets, Stephen Justham, Silvia Toonen and an anonymous referee for very helpful and detailed comments that substantially improved the manuscript.
This research was supported by the Asher Fund for Space Research at the Technion, and the US-Israel Binational Science Foundation. 

% ====================
\section*{APPENDIX \\TIDAL INTERACTION IN WD-CORE BINARIES}
\label{sec:Appendix}
% ====================

We examine the consequence of WD-core tidal interaction in cases of complete envelope ejection. There are very large uncertainties in determining the synchronization time, in particular as the exact convective structure of the exposed
core must be known.
Here we show that even if synchronization is achieved on a short time scale, it will have a small effect on the evolution.

If the core rotation period is longer than the orbital one, the orbital separation will be reduced from $a_f$ to $a_s$.
Assuming the core has a negligible angular momentum before tidal interaction starts, the orbital separation where synchronization is achieved, $a_s$, is given by angular momentum conservation
\begin{equation}
J(a_f)-J(a_s)=I_{\rm core}\omega_{\rm core} (a_s),
\label{eq:J1}
\end{equation}
where $I_{\rm core}$ is the moment of inertia of the core. Substituting for the orbital angular momentum $J$ at $a_f$ and $a_s$, and for the orbital frequency $\omega(a_s)$, equation (\ref{eq:J1}) becomes
\begin{equation}
a_s^{-3/2}I_{\rm core}+a_s^{1/2}\mu=\mu\;a_f^{1/2}
\label{eq:J2}
\end{equation}
Where $\mu$ is the reduced mass of the binary system.
For the typical values in the cases studied here $I_{\rm core} \ll \mu a_f^2$,
and we take the synchronization radius to be
\begin{equation}
a_s=a_f(1-\Delta) \qquad {\rm where} \qquad \Delta \ll 1,
\label{eq:Deltaa1}
\end{equation}
Equation (\ref{eq:J2}) can be cast into the form
\begin{equation}
0 = \frac{I_{\rm core}}{\mu a_f^2}  + (1-\Delta)^2 - (1-\Delta)^{3/2}
\simeq \frac{I_{\rm core}}{\mu a_f^2}  - \frac{1}{2} \Delta,
\label{eq:Deltaa2}
\end{equation}
with the solution
\begin{equation}
\Delta \simeq   \frac{2 I_{\rm core}}{\mu a_f^2}
=  2 \xi \left( \frac{R_{\rm c}} {a_f} \right)^2.
\label{eq:Deltaa3}
\end{equation}
 where
\begin{equation}
\xi \equiv \frac{I_{\rm core}}{\mu R_{\rm core}^2}  .
\label{eq:xi1}
\end{equation}

\begin{figure}[ht]
\centering
\includegraphics[width=80mm]{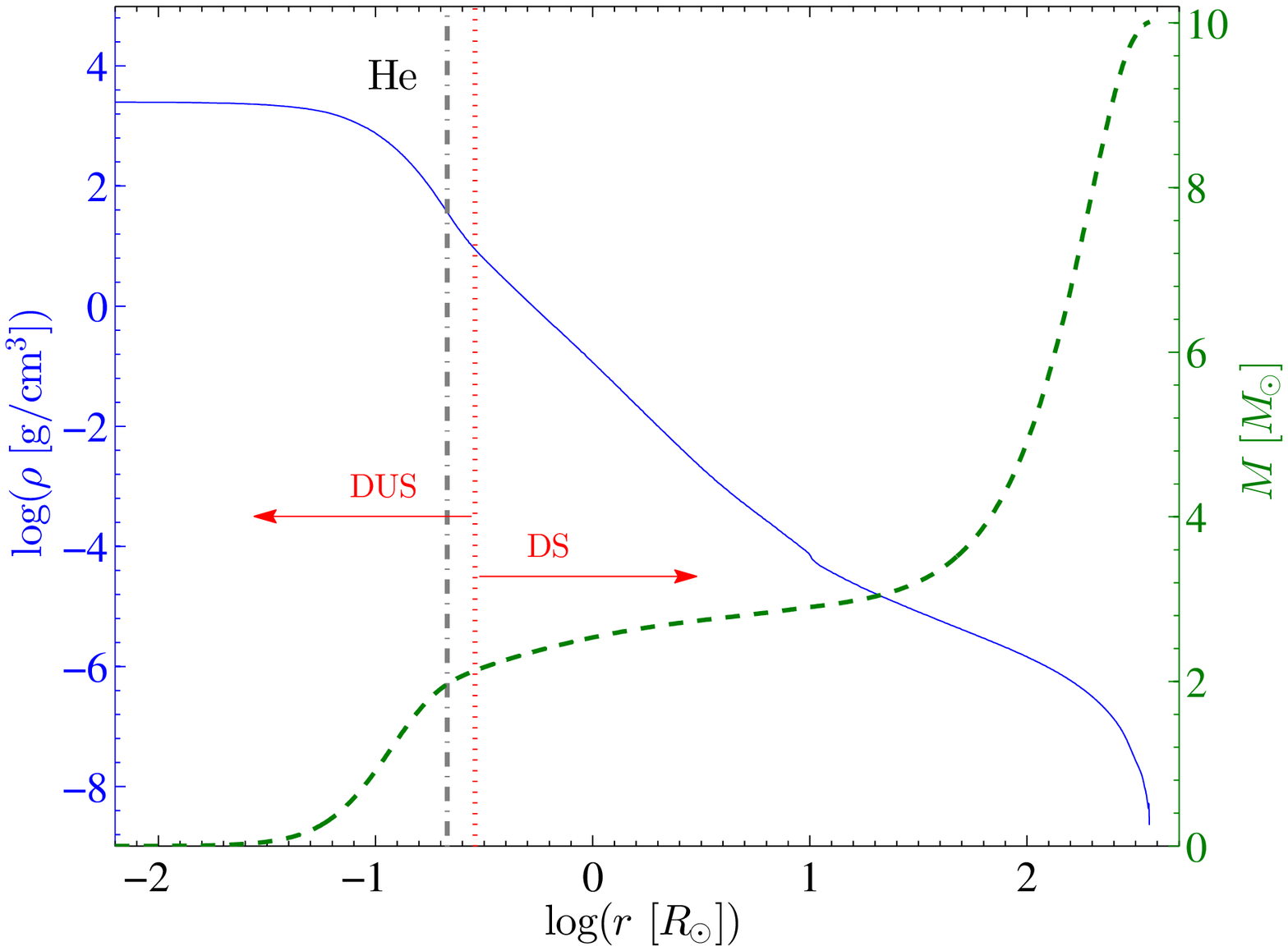}
\hspace*{1mm}
\includegraphics[width=80mm]{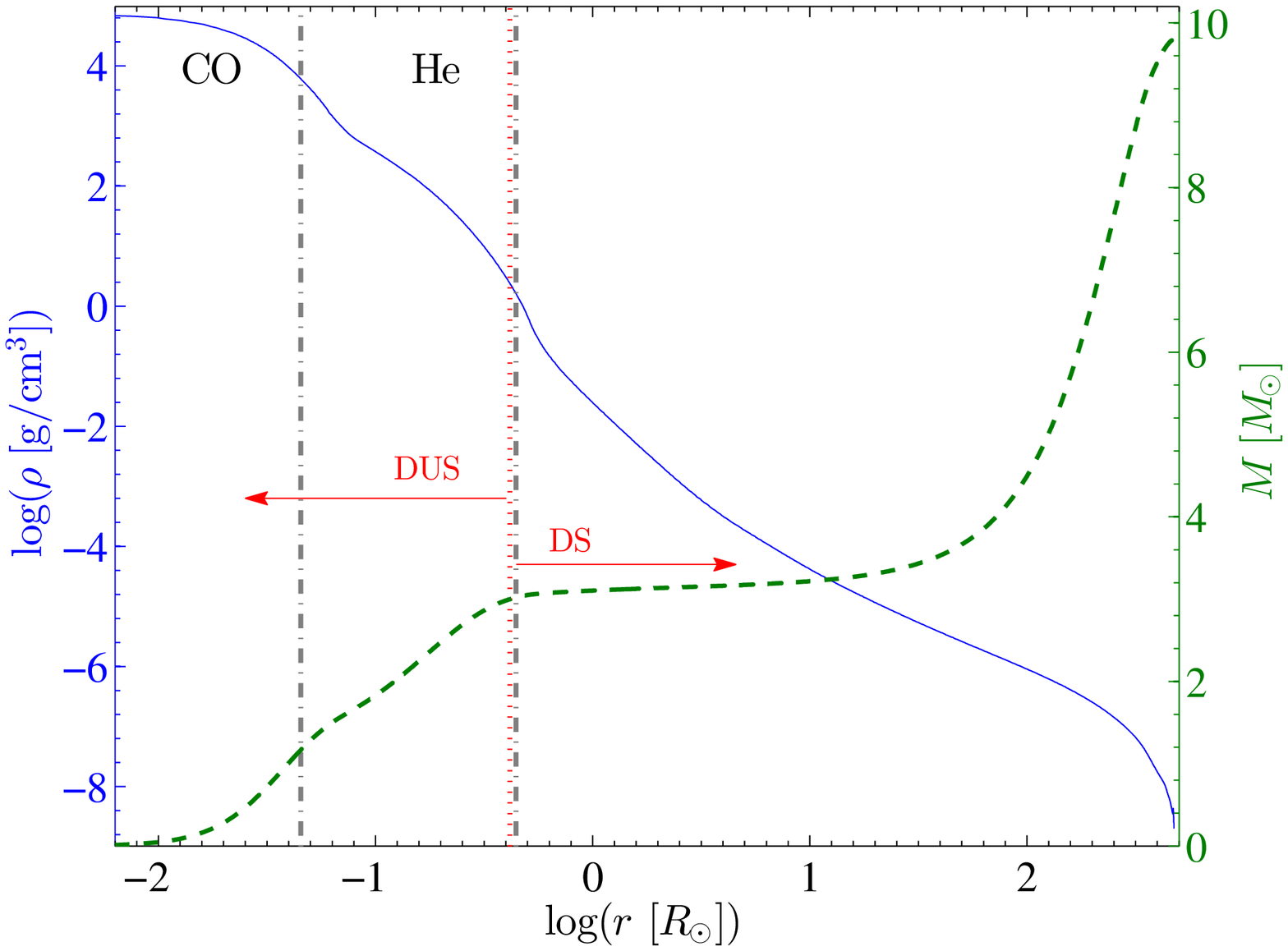}
\hspace*{1mm}
\includegraphics[width=79.5mm]{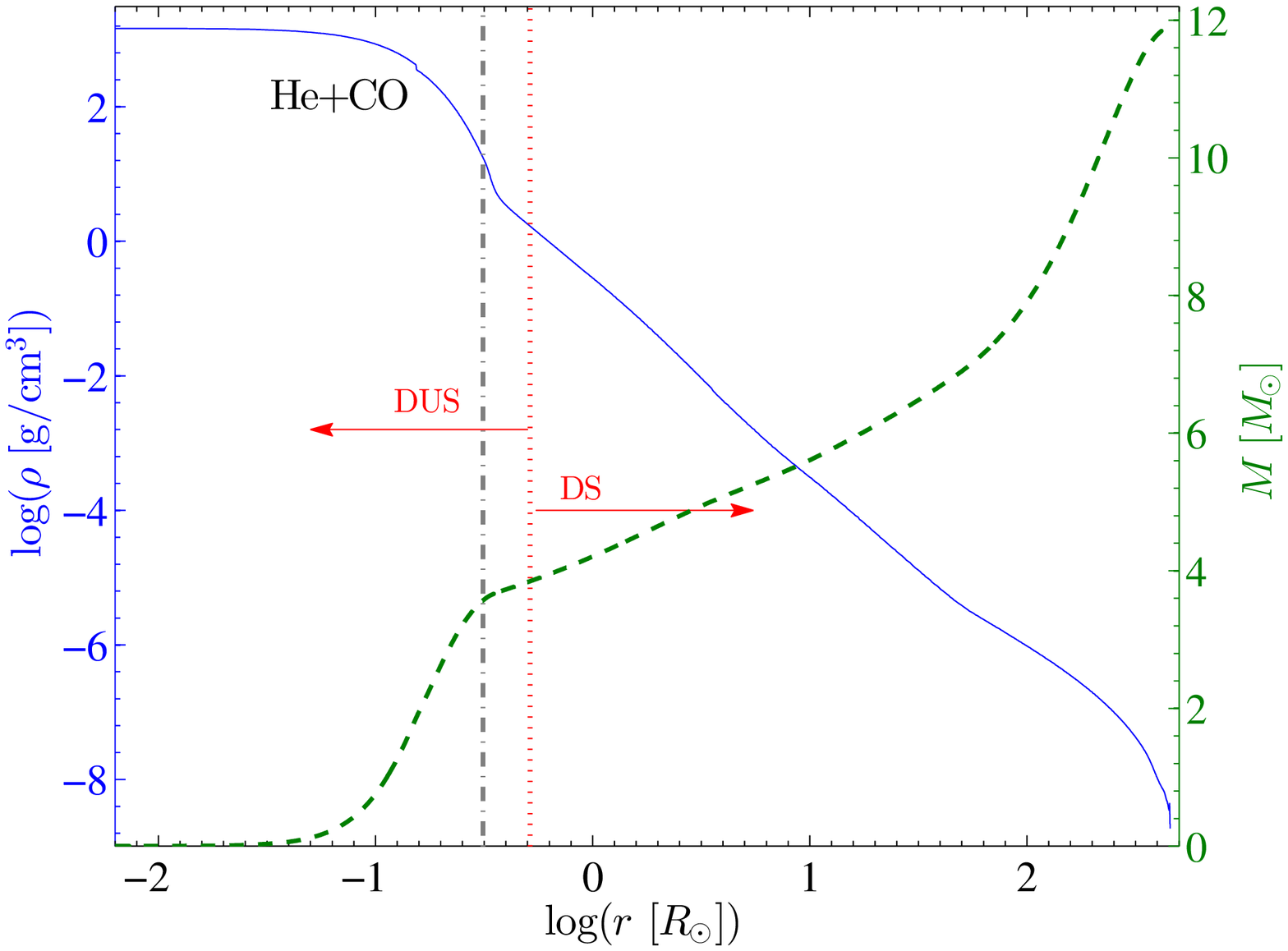}
\hspace*{1mm}
\includegraphics[width=79.5mm]{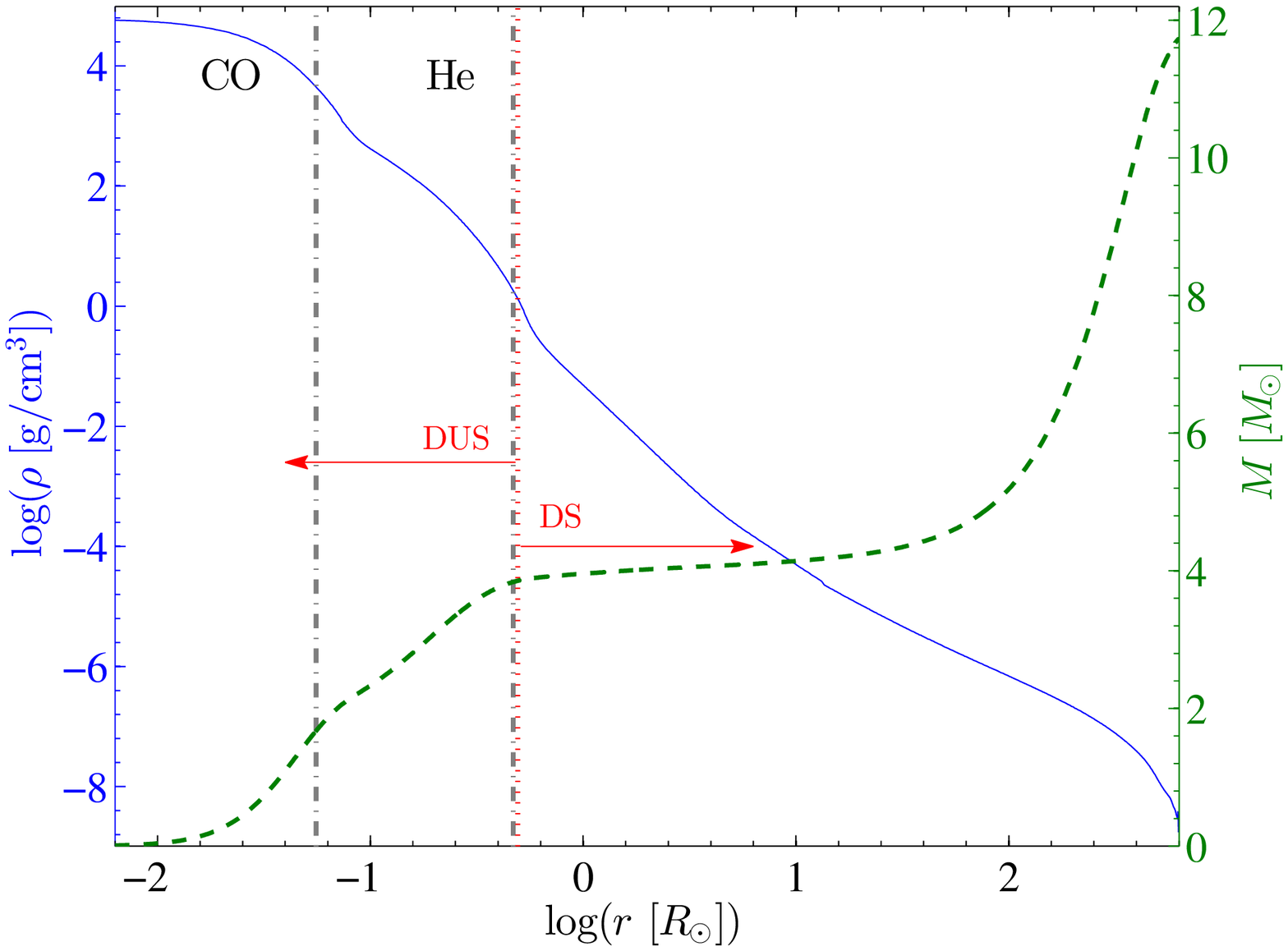}
\caption{Density and stellar mass profiles at the CE phase. The left y-axes relates to the density profile (blue solid line) and the right y-axes relates to the stellar mass profile (thick green dashed line). Upper panels: for a $10M_\odot$ giant with a He core (left panel) and a CO core (right panel). Lower panels: for a $12M_\odot$ giant with a HeCO core (left panel) and a CO core (right panel). The boundary separation at which the system is Darwin unstable is marked with a red dotted line. To the right the system is Darwin stable (DS), to the left it is unstable (DUS).   }
\label{fig:density}
\end{figure}

The values of $\xi$ for the four cases and for $M_{\rm WD}=1 M_\odot$ are given in Table \ref{tab:parameters}.
The distance at which the WD brings the core to synchronization according to equations (\ref{eq:Deltaa1}) and (\ref{eq:Deltaa3}),
 for an initial post-CE orbital separation of $a_f=1 R_\odot$ is given in Table \ref{tab:parameters} as well.
We find that if the orbital separation after envelope ejection is $a_f \ga 1.5 R_{\rm core}$, then the WD does not
spiral-in much while bringing the core to synchronization (tidal locking).
After synchronization is achieved we need to check the stability
of the system against the Darwin instability \citep{Darwin1879}, as also suggested by \cite{Churchetal2006}.
In  a synchronized orbit the Darwin instability sets in when $3 I_{\rm core} > I_{\rm orb}$,
where $I_{\rm orb} = \mu a_s^2$ is the orbital moment of inertia.
The Darwin instability condition reads
\begin{equation}
1< 3 \xi  \left( \frac{R_{\rm core}} {a_f} \right)^2.
\label{eq:Darwin1}
\end{equation}
From the values of $\xi$ given in Table \ref{tab:parameters} we see that for the four cases studied here the system is practically Darwin
stable; only for models (1) \& (3) in Table \ref{tab:parameters} and for $a_s <1.3 R_{\rm core}$ the system is Darwin unstable though very close to the core surface.
This is also evident from Fig. \ref{fig:density}, where the density profiles and envelope mass inward to radius $r$ are given for the models of $10M_\odot$ with He core (upper left) and CO core (upper right), and $12M_\odot$ with HeCO core (lower left) and CO core (lower right).
% FFFFFFFFFFFFFFFFFFFFFFFFFFFFFFFFFFFFFFFFFFFFFFFFFF
% FFFFFFFFFFFFFFFFFFFFFFFFFFFFFFFFFFFFFFFFFFFFFFFFFF

These conclusions, as presented in Table \ref{tab:parameters}, show that tidal interaction, if at all relevant, will have negligible effects on the system and could be disregarded.
From the above discussion we conclude that if the WD manages to eject the entire envelope,
the WD-core binary system will survive at least until further core evolution, or, if there is a fall back gas, a circumbinary disk might cause a merger.
This further evolution is discussed in section \ref{sec:postCE}.

{}


\begin{thebibliography}{}

\bibitem[Arcavi et al.(2012)]{Arcavi2012} Arcavi, I., Gal-Yam, A., Cenko, S.~B., et al.\ 2012, \apjl, 756, L30


\bibitem[Berger et al.(2009)]{Berger2009} Berger, E., et al. 2009, \apj, 699, 1850

\bibitem[Berger et al.(2011)]{Berger2011}Berger, E., Foley, R., \& Soderberg, A. 2011, The Astronomer�s Telegram, 3467

\bibitem[Bond et al.(2009)]{Bond2009} Bond, H.~E., Bedin, L.~R., Bonanos, A.~Z., Humphreys, R.~M., Monard, L.~A.~G.~B., Prieto, J.~L., \& Walter, F.~M.\ 2009, \apjl, 695, L154

\bibitem[Botticella et al.(2009)]{Botticella2009} Botticella, M.~T., et al. 2009, \mnras, 398, 1041

\bibitem[Brown et al.(2001)]{Brownetal2001} Brown, G.~E., Lee, C.-H., Portegies Zwart, S.~F., \& Bethe, H.~A.\ 2001, \apj, 547, 345

\bibitem[Cappellaro et al.(1997)]{Cappellaroetal1997}  Cappellaro, E., Turatto, M., Tsvetkov, D.~Y., et al.\ 1997, \aap, 322, 431

\bibitem[Chevalier(2012)]{Chevalier2012} Chevalier, R.~A.\ 2012, \apjl, 752, L2

\bibitem[Church et al.(2006)]{Churchetal2006} Church, R.~P., Bush, S.~J., Tout, C.~A., \& Davies, M.~B.\ 2006, \mnras, 372, 715

\bibitem[Chomiuk et al.(2011)]{Chomiuk2011} Chomiuk, L., Chornock, R., Soderberg, A.~M., et al.\ 2011, \apj, 743, 114

\bibitem[Dall'Osso et al.(2013)]{Dall'OssoPiran2013} Dall'Osso, S., Piran,
T., \& Shaviv, N.\ 2013, arXiv:1308.0944

\bibitem [Darwin(1879)]{Darwin1879} Darwin G. H., 1879, Proc. R. Soc. Lond., 29, 268

\bibitem[Davies et al.(2002)]{Daviesetal2002} Davies, M.~B., Ritter, H., \& King, A.\ 2002, \mnras, 335, 369
%
\bibitem[Dewi \& Tauris(2000)]{DewiTauris2000} Dewi, J.~D.~M., \& Tauris, T.~M.\ 2000, \aap, 360, 1043 

\bibitem[Ekstrom et al.(2012)]{Ekstrometal2012} Ekstrom, S., Georgy, C., Eggenberger, P., et al.\ 2012, \aap, 537, A146

\bibitem[Gal-Yam(2012)]{Gal-Yam2012} Gal-Yam, A.\ 2012, Science,337, 927

\bibitem[Han et al.(1995)]{Hanetal1995} Han, Z., Podsiadlowski, P.,\& Eggleton, P.~P.\ 1995, \mnras, 272, 800

\bibitem[Iben \& Livio(1993)]{IbenLivio1993} Iben, I., Jr., \& Livio, M.\ 1993, \pasp, 105, 1373

\bibitem[Iben \& Tutukov(1984)]{IbenTutukov1984} Iben, I., Jr., \&
Tutukov, A.~V.\ 1984, \apjs, 54, 335

\bibitem[Ilkov \& Soker(2013)]{IlkovSoker2013} Ilkov, M., \& Soker, N.\ 2013, \mnras, 428, 579

\bibitem[Ivanova(2011)]{Ivanova2011} Ivanova, N.\ 2011, \apj, 730, 76

\bibitem[Ivanova et al.(2013a)]{Ivanovaetal2013a} Ivanova, N., Justham, S., Avendano Nandez, J.~L., \& Lombardi, J.~C.\ 2013a, Science, 339, 433

\bibitem[Ivanova et al.(2013b)]{Ivanovaetal2013b} Ivanova, N., Justham, S., Chen, X., et al.\ 2013b, \aapr, 21, 59

\bibitem[Kalogera et al.(2005)]{Kalogeraetal2005} Kalogera, V., Kim, C., Lorimer, D.~R., Ihm, M., \& Belczynski, K.\ 2005, Binary Radio Pulsars, 328, 261

\bibitem[Kashi et al.(2010)]{Kashietal2010} Kashi, A., Frankowski, A., \& Soker, N.\ 2010, \apjl, 709, L11

\bibitem[Kashi \& Soker(2010a)]{KashiSoker2010a} Kashi, A., \& Soker, N.\ 2010a, \apj, 723, 602

\bibitem[Kashi \& Soker(2010b)]{KashiSoker2010b} Kashi, A., \& Soker, N.\ 2010b, (arXiv:1011.1222)

\bibitem[Kashi \& Soker(2011)]{KashiSoker2011} Kashi, A., \& Soker, N.\ 2011, \mnras, 417, 1466

\bibitem[Kashi et al.(2013)]{Kashietal2013} Kashi, A., Soker, N., \& Moskovitz, N.\ 2013, submitted

\bibitem[Kasliwal et al.(2011)]{Kasliwal2011} Kasliwal, M.~M., et al. 2011, \apj, 730, 134

\bibitem[Kim et al.(2003)]{Kimetal2003} Kim, C., Kalogera, V., \& Lorimer, D.~R.\ 2003, \apj, 584, 985

\bibitem[Kochanek(2011)]{Kochanek2011} Kochanek, C.~S.\ 2011, \apj, 741, 37

\bibitem[Kroupa et al.(1993)]{Kroupa1993} Kroupa, P., Tout, C.~A.,
\& Gilmore, G.\ 1993, \mnras, 262, 545

\bibitem[Kulkarni \& Kasliwal(2009)]{KulkarniKasliwal2009} Kulkarni, S.~R., \& Kasliwal, M.~M.\ 2009, astro2010: The Astronomy and Astrophysics Decadal Survey, 2010, 165

\bibitem[Langer(2012)]{Langer2012} Langer, N.\ 2012, \araa, 50, 107

\bibitem[Mason et al.(2010)]{Mason2010} Mason, E., Diaz, M., Williams, R.~E., Preston, G., \& Bensby, T.\ 2010, \aap, 516, A108

\bibitem[Mould et al.(1990)]{Mould1990} Mould, J., et al. 1990, \apjl, 353, L35

\bibitem[Nelemans et al.(2001)]{Nelemansetal2001} Nelemans, G., Yungelson, L.~R., \& Portegies Zwart, S.~F.\ 2001, \aap, 375, 890

\bibitem[Ofek et al.(2007)]{Ofeketal2007} Ofek, E.~O., Cameron,
P.~B., Kasliwal, M.~M., et al.\ 2007, \apjl, 659, L13

\bibitem[Ofek et al.(2008)]{Ofek2008} Ofek, E.~O., et al. 2008, \apj, 674, 447

\bibitem[Ofek et al.(2013)]{Ofeketal2013} Ofek, E. O. et al.\ 2013, \nat 494, 65

\bibitem[Pakmor et al.(2011)]{Pakmoretal2011} Pakmor, R., Hachinger, S., R{\"o}pke, F.~K., \& Hillebrandt, W.\ 2011, \aap, 528, A117 % Subluminous

\bibitem[Pakmor et al.(2012)]{Pakmoretal2012} Pakmor, R., Kromer, M., Taubenberger, S.,  Sim, S. A., R"opke, F. K., \& Hillebrandt, W.\ 2012, \apjl, 747, L10 % Main

\bibitem[Pastorello et al.(2010)]{Pastorello2010} Pastorello, A., et al. 2010, \mnras, 408, 181

\bibitem[Paxton et al.(2011)]{Paxton2011} Paxton, B., Bildsten,
L., Dotter, A., et al.\ 2011, \apjs, 192, 3

\bibitem[Portegies Zwart \& Verbunt(1996)]{PortegiesZwartVerbunt1996} Portegies Zwart, S.~F., \& Verbunt, F.\ 1996, \aap, 309, 179

\bibitem[Portegies Zwart \& Yungelson(1999)]{PortegiesZwartYungelson1999} Portegies Zwart, S.~F., \& Yungelson, L.~R.\ 1999, \mnras, 309, 26 

\bibitem[Prieto et al.(2009)]{Prieto2009} Prieto, J.~L., Sellgren, K., Thompson, T.~A., \& Kochanek, C.~S.\ 2009, \apj, 705, 1425

\bibitem[Rau et al.(2007)]{Rau2007} Rau, A., Kulkarni, S.~R., Ofek, E.~O., \& Yan, L.\ 2007, \apj, 659, 1536

\bibitem[Raghavan et al.(2010)]{Raghavanetal2010} Raghavan, D., 
McAlister, H.~A., Henry, T.~J., et al.\ 2010, \apjs, 190, 1

\bibitem[Sipior et al.(2004)]{Sipioretal2004} Sipior, M.~S., Portegies
Zwart, S., \& Nelemans, G.\ 2004, \mnras, 354, L49

\bibitem[Smith et al.(2007)]{Smithetal2007} Smith, N., Li, W., Foley,
R.~J., et al.\ 2007, \apj, 666, 1116

\bibitem[Smith et al.(2009)]{Smith2009} Smith, N., et al. 2009, \apjl, 697, L49

\bibitem[Soker(2013a)]{Soker2013a} Soker, N.\ 2013a, \na, 18, 18  % merger in CE

\bibitem[Soker(2013b)]{Soker2013b} Soker, N.\ 2013b, arXiv:1302.5037 % 2012mc

\bibitem[Soker \& Kashi(2013)]{SokerKashi2013} Soker, N., \& Kashi, A.\ 2013, \apjl, 764, L6

\bibitem[Soker \& Tylenda(2003)]{SokerTylenda2003} Soker, N., \& Tylenda, R.\ 2003, \apjl, 582, L105

\bibitem[Tauris \& Dewi(2001)]{TaurisDewi2001} Tauris, T.~M., \& Dewi, J.~D.~M.\ 2001, \aap, 369, 170

\bibitem[Tauris et al.(2013)]{Taurisetal2013} Tauris, T.~M., Sanyal, D., Yoon, S.-C., \& Langer, N.\ 2013, arXiv:1308.4887

\bibitem[Tauris \& Sennels(2000)]{TaurisSennels2000} Tauris, T.~M., \& Sennels, T.\ 2000, \aap, 355, 236

\bibitem[Toonen \& Nelemans(2013)]{ToonenNelemans2013} Toonen, S., \& Nelemans, G.\ 2013, arXiv:1309.0327

\bibitem[Toonen et al.(2012)]{Toonenetal2012} Toonen, S., Nelemans, G., \& Portegies Zwart, S.\ 2012, \aap, 546, A70

\bibitem[Tutukov \& Yungelson(1993)]{TutukovYungelSon1993} Tutukov, A.~V., \& Yungelson, L.~R.\ 1993, Astronomy Reports, 37, 411

\bibitem[Tylenda et al.(2011)]{Tylendaetal2011} Tylenda, R., Hajduk, M., Kami{\'n}ski, T., et al.\ 2011, \aap, 528, A114

\bibitem[Tylenda et al.(2013)]{Tylendaetal2013} Tylenda, R., Kaminski, T., Udalski, A., et al.\ 2013, \aap, 555, A16

\bibitem[van Haaften et al.(2013)]{vanHaaftenetal2013} van Haaften, L.~M., Nelemans, G., Voss, R., et al.\ 2013, \aap, 552, A69

\bibitem[van Kerkwijk \& Kulkarni(1999)]{vanKerkwijkKulkarni1999} van Kerkwijk, M.~H., \& Kulkarni, S.~R.\ 1999, \apjl, 516, L25 


%%%%%%%%%%%%%%%%%%%










\end{thebibliography}
\end{document}